\documentclass[12pt]{article}
\usepackage{amsmath}
\usepackage{amssymb}
\usepackage{amsthm}
\usepackage{graphicx}
\usepackage{subfigure}
\usepackage{changepage}
\usepackage{hyperref}
\usepackage{cite}
\usepackage{url}
\usepackage{breakurl}
\usepackage[small]{caption}
\usepackage{mathtools,xparse,MnSymbol}
\begin{document}
\baselineskip 0.6cm

\def\bra#1{\langle #1 |}
\def\ket#1{| #1 \rangle}
\def\inner#1#2{\langle #1 | #2 \rangle}
\def\brac#1{\llangle #1 \|}
\def\ketc#1{\| #1 \rrangle}
\def\innerc#1#2{\llangle #1 \| #2 \rrangle}
\def\app#1#2{%
  \mathrel{%
    \setbox0=\hbox{$#1\sim$}%
    \setbox2=\hbox{%
      \rlap{\hbox{$#1\propto$}}%
      \lower1.1\ht0\box0%
    }%
    \raise0.25\ht2\box2%
  }%
}
\def\approxprop{\mathpalette\app\relax}
\DeclarePairedDelimiter{\norm}{\lVert}{\rVert}

\begin{titlepage}

\begin{flushright}
\end{flushright}

\vskip 1.2cm

\begin{center}
{\Large \bf Reanalyzing an Evaporating Black Hole}

\vskip 0.7cm

{\large Yasunori Nomura}

\vskip 0.5cm

{\it Berkeley Center for Theoretical Physics, Department of Physics,\\
  University of California, Berkeley, CA 94720, USA}

\vskip 0.1cm

{\it Theoretical Physics Group, Lawrence Berkeley National Laboratory,\\
 Berkeley, CA 94720, USA}

\vskip 0.1cm

{\it Kavli Institute for the Physics and Mathematics of the Universe 
 (WPI),\\
 UTIAS, The University of Tokyo, Kashiwa, Chiba 277-8583, Japan}

\vskip 0.7cm

\abstract{A coherent picture of the quantum mechanics of a 
 collapse-formed, evaporating black hole is presented.  In a 
 distant frame, semiclassical theory in the zone describes 
 microscopic dynamics of only the ``hard modes,'' the modes that 
 are hard enough to be discriminated in the timescale of Hawking 
 emission.  The thermal nature of these modes arises from 
 microcanonical typicality of the full black hole degrees of 
 freedom, mostly composed of the ``soft modes,'' the modes that 
 cannot be discriminated at the semiclassical level.  The hard 
 modes are purified by a combined system of the soft modes and 
 early Hawking radiation, but not by either of them separately. 
 This intrinsically tripartite structure of entanglement is general, 
 regardless of the age of the black hole.  The interior spacetime 
 emerges only at a coarse-grained level.  To describe it, an 
 effective theory can be erected at each time, which applies only 
 to a limited spacetime region determined by the time at which 
 the theory is erected.  The entire interior of the black hole can 
 be described only using multiple effective theories erected at 
 different times, realizing the idea of complementarity.  We analyze 
 implications of the entanglement structure described here for 
 various phenomena, including Hawking evaporation and general 
 information retrieval.  For multiple entangled black holes, 
 it implies that semiclassical objects dropped into different 
 black holes cannot meet in the interior, although each object 
 smoothly enters the horizon of the black hole to which it is 
 falling.  We also discuss physics in Rindler space, elucidating 
 how it is obtained as a smooth limit of the black hole physics.}

\end{center}
\end{titlepage}

\tableofcontents
\newpage

\section{Introduction}
\label{sec:intro}

A black hole is an object in general relativity from which nothing 
can escape.  As in any other object, its entropy, formally defined 
as the logarithm of the number of independent states in a fixed 
energy interval, is infinity at the classical level, $S_{\rm cl} 
= \infty$, which would give zero temperature, $T_{\rm cl} = 
(\partial S_{\rm cl}/\partial E)^{-1} = 0$.  Quantum mechanics, 
however, regulate these to give~\cite{Bekenstein:1973ur,%
Hawking:1974rv,Hawking:1974sw}
\begin{equation}
  S(M) = \frac{{\cal A}(M)}{4 l_{\rm P}^2},
\qquad
  T(M) = \frac{1}{8\pi M l_{\rm P}^2},
\label{eq:BH-intro}
\end{equation}
where $M$ and ${\cal A}(M) = 16\pi M^2 l_{\rm P}^4$ are the mass 
and horizon area of a (Schwarzschild) black hole, respectively, 
$l_{\rm P}$ is the Planck length, and we have adopted natural 
units $c = \hbar = 1$.  A surprising thing is that the entropy 
is proportional to the surface area, rather than the volume, which 
has led to the idea that a fundamental theory of quantum gravity 
is formulated holographically in non-dynamical, lower dimensional 
spacetime~\cite{'tHooft:1993gx,Susskind:1994vu,Bousso:1999xy}.

The fact that a black hole radiates, and so eventually evaporates, 
allows us to understand it as a resonance appearing, e.g., in an 
intermediate stage of a scattering process~\cite{tHooft:1984kcu,%
Page:1993wv}.  (For an extremal black hole, this requires a 
conjectured property of quantum gravity~\cite{ArkaniHamed:2006dz}.) 
In fact, the Anti-de~Sitter (AdS)/Conformal Field Theory (CFT) 
correspondence~\cite{Maldacena:1997re,Gubser:1998bc,Witten:1998qj} 
strongly suggests that formation and evaporation of a black 
hole occurs unitarily, making the concern of information 
loss~\cite{Hawking:1976ra} obsolete.  The unitary evolution 
of a black hole, however, raises another issue of quantum 
information cloning~\cite{JP,Susskind:1993mu}:\ if Hawking 
radiation contains full information about an object that has 
fallen into a black hole earlier, then its simultaneous existence 
with the object in the interior spacetime would imply a cloning 
of quantum information, which is forbidden by linearity of quantum 
mechanics~\cite{Wootters:1982zz}.  An interesting idea addressing 
this issue is called complementarity~\cite{Susskind:1993if}, which 
asserts that information about the interior spacetime is not 
independent of that in Hawking radiation.  The explicit realization 
of this idea, however, has not been clear, and there are even 
arguments that it cannot be implemented consistently with the 
usual postulates of semiclassical physics~\cite{Almheiri:2012rt,%
Almheiri:2013hfa,Marolf:2013dba}.

The purpose of this paper is to reanalyze quantum mechanics of a 
collapse-formed, evaporating black hole, given recent developments. 
We begin with a series of assumptions that seem reasonable and 
are consistent with our latest understanding of the subject, and 
then develop a coherent picture from them.  Essences of the resulting 
picture involve those discussed in Refs.~\cite{Nomura:2014woa,%
Nomura:2014voa,Nomura:2016qum} and Refs.~\cite{Papadodimas:2012aq,%
Papadodimas:2013jku,Papadodimas:2015jra}.  Semiclassical theory 
in a black hole background describes only a small subset of the 
fundamental degrees of freedom, which are distributed nonlocally 
throughout the zone region~\cite{Nomura:2014woa,Nomura:2014voa,%
Nomura:2016qum}.  The interior spacetime appears after 
coarse-graining microscopic degrees of freedom in a state-dependent 
manner~\cite{Papadodimas:2012aq,Papadodimas:2013jku,%
Papadodimas:2015jra}.  The picture also contains an element 
of Ref.~\cite{Maldacena:2013xja} in that the relevant microscopic 
degrees of freedom involve those of early Hawking radiation, 
although the structure of entanglement we find differs from 
that considered there.

While some of the concepts used have been developed in the context 
of AdS/CFT, we will avoid the language of holography as much as 
possible, since the question is mostly about the microscopic picture 
in the bulk.  Our focus will be on Schwarzschild black holes in 
4-dimensional asymptotically flat spacetime (or small black holes 
in 4-dimensional asymptotically AdS spacetime).  However, the 
restriction on specific spacetime dimensions or on non-rotating, 
non-charged black holes is not essential for our discussion.

In the description based on a distant reference frame, our picture 
has the following key features:
\begin{itemize}
\item
In the black hole zone region, semiclassical theory describes 
microscopic dynamics of only the hard modes, the modes that 
are hard enough to be discriminated in the timescale of Hawking 
emission.  These modes comprise only a tiny fraction of the 
total black hole degrees of freedom.  The other, soft modes can 
be described only statistically.
\item
The thermal nature of the hard modes arises because they are in 
equilibrium with the soft modes, the vast majority of the black 
hole degrees of freedom.  In particular, the canonical nature 
of the hard (semiclassical) modes arises from the microcanonical 
ensemble of the full black hole degrees of freedom.
\item
The spatial distribution of the soft modes can be defined 
by interactions with the other modes, and it is given by the 
entropy density determined by the local Hawking temperature 
$T(M)/\sqrt{1-2Ml_{\rm P}^2/r}$.  While this distribution is 
strongly peaked toward the stretched horizon, there are $O(1)$ 
numbers of degrees of freedom located around the edge of the zone. 
Although these comprise only a tiny portion---fractionally of 
$O(l_{\rm P}^2/{\cal A}(M))$---of the full black hole degrees 
of freedom, they play an important role in Hawking emission.
\item
The hard modes in the zone region are purified by a combined system 
of the soft modes and early Hawking radiation, but not by either 
of them separately.  In particular, the correlation between the 
hard modes and either of the soft modes or early Hawking radiation 
is essentially classical.  This is the case regardless of the age 
of the black hole, i.e. if it is younger or older than the Page 
time~\cite{Page:1993wv}.
\end{itemize}
In the distant frame description, the evolution of a black hole is 
unitary, and the interior of the black hole is absent.  In the context 
of holography, this corresponds to the description based on boundary 
time evolution~\cite{Nomura:2018kji}.

The interior spacetime emerges only effectively at a coarse-grained 
level.  The resulting effective theories have the following features:
\begin{itemize}
\item
An effective theory can be erected at each time (of a distant 
description) for the purpose of describing a small object falling 
inside the horizon, until it hits the singularity.
\item
Each effective theory describes only a limited spacetime region 
determined by the time, $t_*$, at which the effective theory is 
erected.  Specifically, the region is the domain of dependence of 
the union of the zone and its mirror regions of a two-sided black 
hole obtained from the original black hole at $t_*$.
\item
The mirror operators needed for an effective theory act on both 
the soft mode and early radiation degrees of freedom.  In particular, 
neither of the soft modes nor early radiation alone can play the 
role of the second exterior of the effective two-sided description.
\item
Since the spacetime region described by each effective theory is 
limited, the entire interior of a black hole can be covered only 
using multiple effective theories erected at different times, 
which are generally not mutually independent.  This provides 
a specific way in which the idea of complementarity is implemented. 
It also provides a simple solution to the cloning paradox that 
no duplicate information occurs in any single description.
\end{itemize}

The entanglement structure between the hard modes, soft modes, 
and early radiation described above is intrinsically tripartite 
and, in a sense, is reminiscent of the Greenberger-Horne-Zeilinger 
(GHZ) form~\cite{GHZ}.  It implies, together with a simple assumption 
about the dynamics of the black hole, that manipulating early 
Hawking radiation alone cannot destroy a smooth horizon of the 
black hole.  It also implies that a pair of entangled Schwarzschild 
(or small AdS) black holes are not connected causally by a wormhole; 
namely, objects dropped into different black holes cannot meet in 
the interior spacetime.  The situation, therefore, is different 
from that in Refs.~\cite{Maldacena:2001kr,Gao:2016bin}, which 
consider entangled large AdS black holes in a thermal state.

In addition to analyzing physics of an evaporating black hole, 
we also consider the Rindler limit.  This elucidates the relation 
between Hawking emission/mining~\cite{Unruh:1982ic,Brown:2012un} 
from a black hole and Unruh radiation~\cite{Unruh:1976db,%
Fulling:1972md,Davies:1974th} seen by an accelerating observer 
in Minkowski space.  In particular, it clarifies ``information 
flow'' associated with the Unruh effect.

\subsubsection*{Relation to other work}

The physics of an evaporating black hole has been studied in 
a large amount of literature, especially after the work of 
Refs.~\cite{Almheiri:2012rt,Almheiri:2013hfa,Marolf:2013dba}, 
some of which have overlaps with the picture presented 
here at conceptual levels.  The fact that the degrees of 
freedom described by semiclassical theory comprise only a tiny 
fraction of the entire degrees of freedom was emphasized in 
Ref.~\cite{Nomura:2013lia}, which was later demonstrated in a 
more convincing form in Ref.~\cite{Almheiri:2014lwa}.  Nonlocality 
associated with the Hawking emission process was considered in 
Refs.~\cite{Giddings:2012bm,Giddings:2012gc,Giddings:2013kcj}, 
although here we do not need a deviation from local dynamics at the 
semiclassical level.  Nonlocality of Hawking emission more similar 
to the one discussed here~\cite{Nomura:2014woa,Nomura:2014voa,%
Nomura:2016qum} was noted in Refs.~\cite{Nomura:2014yka,%
Israel:2015ava,Giddings:2015uzr,Giddings:2017mym}.  State 
dependence of interior operators~\cite{Papadodimas:2012aq,%
Papadodimas:2013jku,Papadodimas:2015jra} was also considered 
in Refs.~\cite{Verlinde:2012cy,Nomura:2012ex,Verlinde:2013uja,%
Nomura:2013gna,Verlinde:2013qya}.  Earlier attempts to avoid 
firewalls along the lines of Ref.~\cite{Maldacena:2013xja} 
include Refs.~\cite{Bousso:2012as,Jacobson:2013ewa,%
Harlow:2013tf,Susskind:2013tg}.  For more recent analyses, 
see Refs.~\cite{deBoer:2018ibj,Hayden:2018khn,Almheiri:2018xdw}.

\subsubsection*{Outline}

In Section~\ref{sec:distant}, we study an evaporating black hole 
as viewed from a distant observer.  In Section~\ref{subsec:micro}, 
we discuss black hole microstates and introduce the concept of the 
hard and soft modes.  In Section~\ref{subsec:static}, we describe 
how the thermal nature of a black hole emerges from a microscopic 
point of view, which elucidates what semiclassical theory is. 
In Section~\ref{subsec:Hawking}, we analyze the Hawking emission 
process.  We discuss how information is transferred from a black 
hole to ambient space, emphasizing that the nonlocal distribution 
of black hole information plays an important role.  We find that 
entanglement between the hard mode, soft mode, and early radiation 
degrees of freedom takes an intrinsically tripartite form, regardless 
of the age of the black hole.

In Section~\ref{sec:interior}, we discuss how the picture of 
the interior spacetime emerges from the microscopic point of 
view.  In Section~\ref{subsec:falling}, we study basic kinematics, 
emphasizing that the equivalence principle dictates the dynamics 
of only a small object, which is well described by the hard modes. 
In Section~\ref{subsec:interior}, we discuss how the effective 
two-sided description may emerge through coarse-graining from 
the entanglement structure discussed in Section~\ref{sec:distant}. 
This allows us to erect an effective theory of the interior at 
each time of a distant description.  While each theory erected 
in this way covers only a limited portion of the interior spacetime, 
the full picture of the interior can be obtained (only) with a 
collection of effective theories.  We also argue that our framework 
provides the ``simplest'' solution to the cloning paradox:\ 
no duplication of information occurs in any single description, 
regardless of whether it can be operationally possessed by an 
observer or not.

In Section~\ref{sec:Rindler}, we discuss the Rindler limit, aiming 
to clarify the relation between Hawking emission/mining from 
a black hole and the Unruh effect in Minkowski space.  In 
Section~\ref{sec:pair}, we consider multiple black holes and 
see that the situation of entangled Schwarzschild (or small 
AdS) black holes is different from that of entangled large 
AdS black holes in a thermal state.  Finally, we conclude in 
Section~\ref{sec:concl}, in which we discuss implications for 
a holographic description and make a few general remarks about 
the black hole interior and singularity.

\section{Black Hole and Information}
\label{sec:distant}

In this section, we discuss how a collapse-formed, or single-sided, 
black hole can be described from the viewpoint of a distant 
observer.  We discuss the interpretation of the Bekenstein-Hawking 
entropy and how the degrees of freedom it represents interact 
with the modes described by semiclassical theory.  We also discuss 
implications of this picture for the Hawking emission process, 
including the evolution of the entanglement structure during 
evaporation.  Throughout, we assume that the evolution of a black 
hole is unitary in a distant description.

\subsection{Black hole microstates}
\label{subsec:micro}

Consider a set of states having energies, as measured in the 
asymptotic region, between $E$ and $E + \varDelta E$.  Some of 
these states can be recognized from the asymptotic region as those 
representing multiple particle excitations of masses $m_i$:\ 
$\sum_i m_i \approx E$.  Such a decomposition is possible if these 
particles are distributed with sufficiently large distances between 
them.  There are, however, states in which this decomposition cannot 
be made completely.  These include states having black holes.  (They 
also include states having coherent excitations.)  For simplicity, 
we will focus on states that have a single Schwarzschild black hole.

Let us consider a quantum state representing a black hole of mass 
$M$ located at some place at rest, where the position and velocity 
are measured with respect to a distant reference frame.  Because of 
the uncertainty principle, such a state must involve a superposition 
of energy and momentum eigenstates.  According to the standard Hawking 
calculation, a state of a black hole of mass $M$ will evolve after 
Schwarzschild time $t_{\rm H} \approx O(M l_{\rm P}^2)$ into a state 
representing a Hawking quantum of energy $\approx O(1/M l_{\rm P}^2)$ 
and a black hole with the correspondingly smaller mass.%
\footnote{Note that despite the apparent language here, the evolution 
 of the state is continuous.  Specifically, when the state is 
 expanded in the eigenbasis of Hawking quanta, dominant terms 
 of the state shift continuously to those with one more Hawking 
 quantum with the characteristic timescale of $O(M l_{\rm P}^2)$.}
The fact that these two states---before and after the emission---are 
nearly orthogonal implies that the original state must involve a 
superposition of energy eigenstates with a spread, at least, of
\begin{equation}
  \varDelta E \approx \frac{1}{t_{\rm H}} 
    \approx O\biggl(\frac{1}{M l_{\rm P}^2}\biggr).
\label{eq:delta-E}
\end{equation}
Black hole states within this energy range cannot be discriminated 
in the asymptotic region because such a discrimination would require 
time longer than $t_{\rm H}$, the timescale with which a black hole 
state changes to another, orthogonal state.  These states, therefore, 
comprise the {\it microstates} of a black hole of mass $M$.

What about the spread of momentum $\varDelta p$, with $p$ measured 
in the asymptotic region?  Let us assume that the spatial location 
of the black hole is identified with precision comparable to the 
quantum stretching of the horizon $\varDelta d \approx O(l_{\rm s})$, 
namely $\varDelta r \approx O(l_{\rm s}^2/M l_{\rm P}^2)$, where 
$d$ and $r$ are the proper length and the Schwarzschild radial 
coordinate, respectively, and $l_{\rm s}$ is the string (cutoff) 
length.  This implies that a black hole state must involve a 
superposition of momentum eigenstates with spread $\varDelta p 
\approx (l_{\rm s}/M l_{\rm P}^2) (1/\varDelta d) \approx 
O(1/M l_{\rm P}^2)$.  Here, the factor $l_{\rm s}/M l_{\rm P}^2$ 
in the middle expression is the redshift factor.  This value 
of $\varDelta p$ corresponds to an uncertainty of the kinetic 
energy $\varDelta E_{\rm kin} \approx (\varDelta p)^2/M \approx 
O(1/M^3 l_{\rm P}^4)$, which is much smaller than $\varDelta E$ 
in Eq.~(\ref{eq:delta-E}).  The spread of energy, therefore, 
comes mostly from a superposition of different rest masses:\ 
$\varDelta E \approx \varDelta M$.

The number of microstates for a black hole is given by the 
Bekenstein-Hawking formula.  Specifically,  the number of independent 
microstates, ${\cal N}(M)$, for a black hole of mass $M$ is given by
\begin{equation}
  {\cal N}(M) = \exp \biggl[\frac{{\cal A}(M)}{4 l_{\rm P}^2}\biggr] 
    \frac{\varDelta M}{M} 
  \equiv e^{S_{\rm BH}(M)} \frac{\varDelta M}{M},
\label{eq:S_BH}
\end{equation}
where
\begin{equation}
  {\cal A}(M) = 16\pi M^2 l_{\rm P}^4,
\label{eq:A_M}
\end{equation}
is the area of the black hole horizon.  The width of the mass 
range, $\varDelta M$, is given by Eq.~(\ref{eq:delta-E}), although 
the value of the Bekenstein-Hawking entropy, $S_{\rm BH}(M)$, is 
insensitive to the precise choice of $\varDelta M$, as in usual 
statistical mechanical systems.

The discussion above implies that it is not appropriate to consider 
that quantum mechanics introduces exponentially large degeneracies 
for black hole microstates which did not exist in a classical black 
hole.  In classical general relativity, a set of Schwarzschild 
black holes located at some place at rest are parameterized by 
a continuous mass parameter $M$; i.e., there are a continuously 
infinite number of black hole states in the energy interval between 
$M$ and $M + \varDelta M$ for any $M$ and small $\varDelta M$. 
Quantum mechanics reduces this to a finite number $\approx 
e^{S_{\rm BH}(M)} \varDelta M/M$.%
\footnote{Of course, quantum mechanics allows for a superposition of 
 these finite number of independent states, so the number of possible 
 (not necessarily independent) states is continuously infinite.  The 
 statement here applies to the number of independent states, regarding 
 classical black holes with different $M$ as independent states.}
This can also be seen from the fact that $S_{\rm BH}(M)$ is written 
as ${\cal A}(M) c^3/4 \hbar l_{\rm P}^2$ when $c$ and $\hbar$ are 
restored, which becomes infinite for $\hbar \rightarrow 0$ with $c$ 
and $l_{\rm P}$ fixed.  Indeed, this situation is quite standard in 
the relation between quantum and classical mechanics.  For example, 
the number of independent states of a harmonic oscillator in a fixed 
energy interval is finite in quantum mechanics (labeled by a discrete 
number for the levels) while it is infinite in classical mechanics 
(labeled by a continuous amplitude).

\subsection{Semiclassical description and the static background 
 approximation}
\label{subsec:static}

From now on, we will suppress the location of a black hole and write 
a quantum state containing a black hole of mass $M$ as
\begin{equation}
  \ket{\Psi(M)} \approx \ket{\psi(M)} \ket{\phi},
\label{eq:state}
\end{equation}
where $\ket{\psi(M)}$ represents the state of the system within 
the zone region, $r \leq r_{\rm z} \approx 3 M l_{\rm P}^2$, while 
$\ket{\phi}$ represents the state of the far region, $r > r_{\rm z}$. 
As discussed in the previous subsection, the mass $M$ is specified 
with precision $\varDelta M \approx 1/M l_{\rm P}^2$.  The 
separation of the state as in Eq.~(\ref{eq:state}) is justified 
because the degrees of freedom associated with the black hole 
microstates---represented by the thermal atmosphere of the black 
hole in the semiclassical description---are confined in the region 
$r \leq r_{\rm z}$.%
\footnote{Strictly speaking, $\ket{\phi}$ also depends on $M$, 
 but we suppress this argument because it is not important for 
 our purposes.  We will also treat excitations spreading both in 
 the $r \leq r_{\rm z}$ and $r > r_{\rm z}$ regions approximately 
 by including them either in $\ket{\psi(M)}$ or $\ket{\phi}$. 
 The precise description of these excitations will require a 
 more elaborate expression, but we believe this is an inessential 
 technical subtlety in addressing our problem.}

When sufficient time is passed after a black hole is formed by 
collapse, the state of the entire system is given by a superposition 
of terms of the form in Eq.~(\ref{eq:state}) with different black 
hole masses and locations.  The superposition necessarily arises 
because of the backreaction of Hawking emission~\cite{Page:1979tc,%
Nomura:2012cx}.  The full unitarity of time evolution is retained 
only if we keep all these terms.  However, including this effect is 
straightforward, and it provides only minor corrections to entropic 
considerations.  We thus use the form of Eq.~(\ref{eq:state}) in 
discussing the dynamics of the black hole.

After being equilibrated, a black hole can be viewed as static in 
a timescale shorter than $t_{\rm H} \approx M l_{\rm P}^2$, at least 
semiclassically.  How does the black hole state $\ket{\psi(M)}$ look 
then?  For now, we ignore the state $\ket{\phi}$ when discussing the 
black hole.  As we will see in Section~\ref{subsubsec:micro}, this 
assumption is not quite justified, except possibly for some very 
early time, but it serves as a good starting point for discussion. 
Let us consider that the black hole state is perturbed by 
excitations, e.g.\ by an infalling object.  Such an object 
can be well described using modes whose frequency $\omega$ 
(as measured in the asymptotic region) is sufficiently larger 
than the Hawking temperature
\begin{equation}
  T_{\rm H} = \frac{1}{8\pi M l_{\rm P}^2},
\label{eq:T_H}
\end{equation}
as we will see more explicitly in Section~\ref{subsec:falling}. 
We therefore separate out these high frequency modes from the rest 
and call them the {\it hard modes}.  Below, we write the condition 
for the hard modes as $\omega \geq \Delta \gg T_{\rm H}$, although 
$\Delta$ is not too large:\ $\Delta \approx O(10)T_{\rm H} \approx 
O(1)/M l_{\rm P}^2$.

We group all the orthonormal black hole microstates (in some basis) 
into sets in such a way that states in the same set have the same 
configuration for the hard modes.  We can do this approximately, 
which is enough.  By labeling the sets by the energy, $E$, carried 
by the hard modes and their members by $i_E$, a black hole microstate 
can be written as
\begin{equation}
  \ket{\psi(M)} = \sum_E \sum_{i_E = 1}^{{\cal N}(M-E)} 
      c_{E i_E} \ket{E} \ket{\psi_{i_E}(M-E)};
\qquad
  \sum_E \sum_{i_E = 1}^{{\cal N}(M-E)} |c_{E i_E}|^2 = 1.
\label{eq:BH-states}
\end{equation}
Here, $\ket{E}$ are orthonormal states of the hard modes.  
$\ket{\psi_{i_E}(M-E)}$ for each $E$ are orthonormal states 
representing members of the set, and we call the modes associated 
with these states the {\it soft modes}.  ${\cal N}(M-E)$ is given 
by Eq.~(\ref{eq:S_BH}), where we have identified $\varDelta M$ 
and $\Delta$; as mentioned at the end of the paragraph containing 
Eq.~(\ref{eq:S_BH}), this does not cause any error in the 
statistical limit.  Note that the Hilbert spaces for the hard 
and soft modes, as defined here, do not factorize because of 
the energy constraint.

In writing Eq.~(\ref{eq:BH-states}), we have assumed that 
the number of independent states for the hard modes is much 
smaller than that of the soft modes.  We will discuss this in 
Section~\ref{subsubsec:micro}; here we merely point out that we 
only consider states that do not yield a significant backreaction 
on spacetime, which limits the number of possible hard mode states.%
\footnote{This implies that the algebra defined on the space of 
 $\ket{E}$ does not close, the situation that often appears in 
 quantum gravity.  Here we treat the space as a simple Hilbert 
 space; for a mathematically more rigorous treatment, see, 
 e.g., Ref.~\cite{Ghosh:2017gtw}.}
We also assume that both types of modes exist only in the region 
outside the stretched horizon, $r \geq r_{\rm s}$, where
\begin{equation}
  r_{\rm s} - 2 M l_{\rm P}^2 
  \approx O\biggl(\frac{l_{\rm s}^2}{M l_{\rm P}^2}\biggr).
\label{eq:stretched}
\end{equation}
This is motivated by the fact that the spacetime picture breaks 
down in the region $r < r_{\rm s}$ due to stringy effects, and that 
the Bekenstein-Hawking entropy is reproduced by integrating the 
entropy density of the thermal atmosphere in the region $r_{\rm s} 
\leq r \leq r_{\rm z}$, as we will see below.  We note that for 
$\ket{E}$ we include modes on the stretched horizon, e.g.\ its 
vibration modes.

Our central assertion is that it is only the hard modes that 
semiclassical theory can describe at the full quantum level.  Here, 
by semiclassical theory we mean quantum theory with gravity defined 
on a curved spacetime background (see Section~\ref{subsec:falling} 
for further discussion).  This makes sense because the other, 
soft modes cannot be discriminated in the asymptotic region 
within a timescale at which the black hole background can be 
viewed as static.  This implies that the black hole state at 
the semiclassical level is obtained after tracing out these soft 
modes.  This leads to
\begin{align}
  \rho(M) &= \sum_E \sum_{i_E = 1}^{{\cal N}(M-E)} 
    |c_{E i_E}|^2 \ket{E} \bra{E} 
\nonumber\\
  &\simeq \frac{1}{\sum_E {\cal N}(M-E)} 
    \sum_E {\cal N}(M-E) \ket{E} \bra{E} 
\nonumber\\
  &= \frac{1}{\sum_E e^{-\frac{E}{T_{\rm H}}}} 
    \sum_E e^{-\frac{E}{T_{\rm H}}} \ket{E} \bra{E}.
\label{eq:BH-thermal}
\end{align}
To go to the second line, we have assumed that the black hole state 
is generic, i.e.\ the values of $|c_{E i_E}|^2$ are statistically 
the same,%
\footnote{If we consider the microcanonical ensemble of the black 
 hole microstates, then we obtain the expression in the second 
 line directly.}
and in the last expression we have taken the statistical limit 
$E \ll M$, which we will denote by equality.  The expression 
in Eq.~(\ref{eq:BH-thermal}) appears as the standard black 
hole thermal state describing the region $r_{\rm s} \leq r \leq 
r_{\rm z}$.  An important point, however, is that the states 
$\ket{E}$ are supposed to represent only those of the hard 
modes---semiclassical theory does not allow us to describe 
the microscopic dynamics of the modes associated with energy 
differences smaller than $\Delta$.  The consistency of this 
picture will be discussed throughout the paper.

\subsection{Hawking emission}
\label{subsec:Hawking}

In this subsection, we will consider implications of the above 
picture for Hawking emission and evaporation.  The discussion 
below follows initially that of Refs.~\cite{Nomura:2014woa,%
Nomura:2014voa,Nomura:2016qum}, adjusted to the current context. 
The arguments toward the end of Section~\ref{subsubsec:transf} 
and in Section~\ref{subsubsec:micro} are new.

\subsubsection{Distribution of microscopic information}
\label{subsubsec:distr}

In the present picture, semiclassical theory can describe the 
microscopic dynamics of only the hard modes---the soft modes can 
be characterized only as a thermal bath of temperature $T_{\rm H}$ 
(without a hard component), with which the hard modes interact. 
It is, however, not only the hard modes in $r_{\rm s} \leq r \leq 
r_{\rm z}$ that can interact with the soft modes.  Some of the 
modes described by $\ket{\phi}$ in Eq.~(\ref{eq:state}), i.e.\ 
the ``far modes'' in $r > r_{\rm z}$, can also interact with (a 
small fraction of) the soft modes.

To understand this, let us discuss the spatial distribution of the 
soft modes, represented by $\ket{\psi_{i_E}(M-E)}$.  The concept 
of spatial distribution for the soft modes is meaningful despite 
the fact that their internal dynamics is not known.  It is 
defined through interactions with the hard and far modes, which 
we will call the ``semiclassical modes.''  Since the dynamics 
of the semiclassical modes are described by semiclassical theory, 
there is a well-defined notion of where these modes are located. 
The distribution of the soft modes can then be determined 
by analyzing which of the semiclassical modes they mostly 
interact with.

The expression in Eq.~(\ref{eq:BH-thermal}) is consistent with the 
interpretation that the soft and hard modes form an almost closed 
system equilibrated at the temperature $T_{\rm H}$.  It is, 
therefore, reasonable to assume that from the viewpoint of the 
semiclassical modes, the distribution of the soft modes is given 
by the thermal entropy density of a system with temperature 
$T_{\rm H}$ as measured in the asymptotic region.  (The deviation 
from it due to the lack of the hard component is negligible.) 
Since the black hole microstates comprise the soft modes of all 
low-energy species, this implies that the spatial distribution 
of black hole information is given by the entropy density
\begin{equation}
  s(r) = c N(r)\, T_{\rm loc}(r)^3;
\qquad
  T_{\rm loc}(r) = \frac{T_{\rm H}}{\sqrt{1-\frac{2M l_{\rm P}^2}{r}}},
\label{eq:s-density}
\end{equation}
where $c$ is a constant of $O(1)$, $T_{\rm loc}(r)$ is the local 
temperature measured at $r$, and $N(r)$ is the number of low-energy 
species existing below $T_{\rm loc}(r)$.  We find that integrating 
this density over the region $r_{\rm s} \leq r \leq r_{\rm z}$ 
indeed reproduces the Bekenstein-Hawking entropy up to an 
incalculable $O(1)$ factor:
\begin{equation}
  \int_{r_{\rm s}}^{r_{\rm z}}\! s(r)\, 
    \frac{r^2 dr d\Omega}{\sqrt{1-\frac{2M l_{\rm P}^2}{r}}} 
  \approx N(r_{\rm s}) \frac{M^2 l_{\rm P}^4}{l_{\rm s}^2} 
  \approx S_{\rm BH}(M),
\label{eq:int-q}
\end{equation}
where we have assumed that the change of $N(r)$ is not too rapid as 
a function of $r$ and used the relation expected in any theory of 
quantum gravity (see, e.g., Ref.~\cite{Dvali:2007hz}):
\begin{equation}
  l_{\rm P}^2 \approx \frac{l_{\rm s}^2}{N(r_{\rm s})}.
\label{eq:qg-rel}
\end{equation}
Note that if we take the lower limit of the integral in 
Eq.~(\ref{eq:int-q}) to be the classical horizon, $r_{\rm s} 
\rightarrow 2M l_{\rm P}^2$, then the integral diverges.  This 
is consistent with the fact that the entropy of a black hole is 
infinite at the classical level.

To elucidate the significance of this result in our context, we go 
to the tortoise coordinate
\begin{equation}
  r^* = r + 2M l_{\rm P}^2\, 
    \ln\frac{r-2M l_{\rm P}^2}{2M l_{\rm P}^2},
\label{eq:tortoise}
\end{equation}
in which the region outside the Schwarzschild horizon $r \in 
(2M l_{\rm P}^2, \infty)$ is mapped into $r^* \in (-\infty,\infty)$. 
This coordinate is useful in that the kinetic term of an appropriately 
redefined field takes the canonical form, so that its propagation can 
be analyzed as in flat space.  In this coordinate, the stretched 
horizon, located at $r_{\rm s} = 2M l_{\rm P}^2 + O(l_{\rm s}^2/M 
l_{\rm P}^2)$, is at
\begin{equation}
  r^*_{\rm s} \simeq -4M l_{\rm P}^2 
    \ln\frac{M l_{\rm P}^2}{l_{\rm s}} 
  \simeq -4M l_{\rm P}^2 \ln(M l_{\rm P}),
\label{eq:rs_stretched}
\end{equation}
where we have taken $l_{\rm s}$ to be not too far from $l_{\rm P}$. 
This implies that there is a large distance between the stretched 
horizon and the potential barrier region around the edge of the 
zone when measured in $r^*$:\ $\varDelta r^* \simeq 4M l_{\rm P}^2 
\ln(M l_{\rm P}) \gg O(M l_{\rm P}^2)$ for $\ln(M l_{\rm P}) \gg 1$. 
On the other hand, a localized Hawking quantum is represented by a 
wavepacket with width of $O(M l_{\rm P}^2)$ in $r^*$, since it has 
an energy of order $T_{\rm H} = 1/8\pi M l_{\rm P}^2$ defined in 
the asymptotic region.

An important point is that the amount of integrated entropy 
contained around the edge of the zone is of $O(1)$:
\begin{equation}
  \int_{|r^*| \lesssim O(M l_{\rm P}^2)}\! s(r(r^*))\, 
    \sqrt{1-\frac{2M l_{\rm P}^2}{r(r^*)}}\, r^2(r^*)\, dr^* d\Omega
  \approx O(1).
\label{eq:entropy}
\end{equation}
While this is a negligibly small fraction of the total black 
hole entropy, of order $l_{\rm P}^2/{\cal A}(M) \ll 1$, it 
has a significant implication for the interpretation of the 
Hawking emission process.  It implies that outgoing field theory 
modes---specifically, outgoing modes represented by $\ket{\phi}$ 
in Eq.~(\ref{eq:state}) and located around $r \approx r_{\rm z}$ 
at the relevant time---can extract black hole information directly 
from the soft modes at the edge of the zone, without involving 
a semiclassical mode deep in the zone.  In other words, from the 
viewpoint of the semiclassical modes, microscopic information 
about the black hole is delocalized over the entire zone, although 
the distribution is strongly peaked toward the stretched horizon.

While the microscopic dynamics of this information extraction 
process cannot be described within semiclassical theory 
because it involves soft modes, the flow of energy and entropy 
can be investigated using energy-momentum conservation, 
thermodynamic considerations, and unitarity.  This is done 
below in Section~\ref{subsubsec:transf}, resulting in the 
following picture.  A Hawking quantum that can be viewed 
as a semiclassical mode is emitted at the edge of the zone, 
where it extracts $O(1)$ information from the soft modes in 
each emission timescale of $t \approx O(M l_{\rm P}^2)$.  Since 
Hawking evaporation is a long process, this small rate is enough 
for all the black hole information to be returned to ambient 
space in the lifetime of the black hole, $\tau_{\rm BH} \approx 
O(M^3 l_{\rm P}^4)$.

\subsubsection{Information transfer}
\label{subsubsec:transf}

Let us now amplify the discussion above using a qubit model.  Take 
the black hole state in Eq.~(\ref{eq:BH-states}).  We focus on the 
terms with the lowest $E$ because for $\Delta \gg T_{\rm H}$ they 
statistically dominate the process:
\begin{equation}
  \ket{\psi(M)} \approx \sum_{i = 1}^{{\cal N}(M)} 
      c_i \ket{E_0} \ket{\psi_i(M)},
\label{eq:BH-states-0}
\end{equation}
where $E_0 = 0$ is the lowest value of $E$, and $\Delta$ is the 
energy separating the hard and soft modes.  In the timescale 
of $O(M l_{\rm P}^2)$, this state changes by emitting a Hawking 
quantum.%
\footnote{Throughout this paper, we assume that the number of species 
 below $T_{\rm H}$ is small, $N(r_{\rm z}) \approx O(1)$, and we 
 mostly focus on the case with a single species.  Including the 
 effect of multiple species is straightforward.}
Suppose that the black hole releases 1~qubit of information 
through Hawking emission.  The energy of the emitted quantum 
is then $\varDelta M \simeq (\ln 2)/8\pi M l_{\rm P}^2$, so that 
${\cal N}(M-\varDelta M) = {\cal N}(M)/2$.  We can model this 
process by saying that the emitted Hawking quantum is in states 
$\ket{r_1}$ and $\ket{r_2}$ if the index for the soft mode states, 
$i$, is odd and even, respectively.%
\footnote{In this model, the qubit that escapes from the black 
 hole is the odd-even direction of $i$ in the space spanned by 
 $\ket{\psi_i(M)}$.  This corresponds to the statement that the 
 information leaving the black hole is that associated with the 
 soft mode located around $r \sim r_{\rm z}$.}
Because of energy-momentum conservation, the process is accompanied 
by the creation of an ingoing negative energy excitation, which 
we denote by a star; namely, $\ket{\psi^*_i(M)}$ represents the 
states of the soft modes with the negative energy excitation.

One might naively think that this process simply goes as
\begin{equation}
  \ket{E_0} \ket{\psi_i(M)} \ket{\phi} 
  \stackrel{?}{\rightarrow} \left\{ \begin{array}{ll}
    \ket{E_0} \ket{\psi^*_i(M)} \ket{\phi + r_1} & 
      \mbox{if $i$ is odd}, \\
    \ket{E_0} \ket{\psi^*_i(M)} \ket{\phi + r_2} &
      \mbox{if $i$ is even},
  \end{array} \right.
\label{eq:toy-1}
\end{equation}
where $\ket{\phi + r_a}$ ($a=1,2$) is the state in which the outgoing 
Hawking quantum in state $\ket{r_a}$ is added to the far state 
$\ket{\phi}$ around the edge of the zone (with the appropriate 
time evolution).  However, this leads to a problem.  Remember that 
$\ket{\psi^*_i(M)}$ have energy $M - \varDelta M$, and we expect 
that they will relax into states of the black hole of the decreased 
mass $M - \varDelta M$:
\begin{equation}
  \ket{\psi^*_i(M)} \rightarrow \ket{\psi_{i'}(M-\varDelta M)}.
\label{eq:toy-2}
\end{equation}
Since $i'$ runs only over $i' = 1,\cdots,{\cal N}(M-\varDelta M) 
= {\cal N}(M)/2$, however, such relaxation cannot occur unitarily. 
Instead, what happens in the emission process must be like
\begin{equation}
  \ket{E_0} \ket{\psi_i(M)} \ket{\phi} 
  \rightarrow \left\{ \begin{array}{ll}
    \ket{E_0} \ket{\psi^*_{\frac{i+1}{2}}(M)} \ket{\phi + r_1} & 
      \mbox{if $i$ is odd}, \\
    \ket{E_0} \ket{\psi^*_{\frac{i}{2}}(M)} \ket{\phi + r_2} &
      \mbox{if $i$ is even},
  \end{array} \right.
\label{eq:toy-3}
\end{equation}
i.e.\ the index for the soft mode states with the negative energy 
excitation runs only from $1$ to ${\cal N}(M)/2$.  This allows 
for these states to relax unitarily into the unexcited soft 
mode states with the decreased mass $M - \varDelta M$, as in 
Eq.~(\ref{eq:toy-2}).  Note that the process in Eq.~(\ref{eq:toy-3}) 
itself is also unitary if we consider the whole quantum state, 
including both the black hole and far regions.

The above analysis says that a negative energy excitation over 
static black hole states, corresponding to the Hartle-Hawking 
vacuum~\cite{Hartle:1976tp} at the semiclassical level, carries 
a negative entropy.  Namely, in the existence of a negative energy 
excitation, the range over which the microstate index $i$ runs is 
smaller than that without.  This shows that the standard relation 
between entropy and energy, $S \sim E$, persists even if these 
quantities are defined with respect to a static black hole 
background.  Specifically, the excitation of energy $-\varDelta M$ 
carries entropy
\begin{equation}
  \varDelta S = -8\pi M \varDelta M l_{\rm P}^2 
  = \frac{-\varDelta M}{T_{\rm H}}.
\label{eq:neg-S}
\end{equation}
Since a negative energy excitation does not relax instantaneously, 
the initial states in Eq.~(\ref{eq:toy-3}) may contain multiple 
negative energy excitations created by earlier emissions.  However, 
we expect that the relaxation time of a negative energy excitation 
is not much larger than $O(M l_{\rm P}^2 \ln(M l_{\rm P}))$, the 
time it takes for an excitation to propagate from the edge of 
the zone to the stretched horizon and also the time it takes for 
information to be scrambled~\cite{Hayden:2007cs,Sekino:2008he}. 
Thus, the number of negative energy excitations existing at any 
moment is expected to be $\lesssim O(\ln(M l_{\rm P}))$, which is 
exponentially smaller than that of the degrees of freedom 
represented by $i$.  This implies that negative energy excitations 
created earlier lie in regions far from the edge of the zone, 
and hence their effects on the process of Eq.~(\ref{eq:toy-3}) 
can be safely ignored.%
\footnote{We may redefine the semiclassical vacuum by including 
 these negative energy-entropy excitations.  The resulting 
 vacuum will correspond, very roughly, to the Unruh 
 vacuum~\cite{Unruh:1976db}, and the associated geometry 
 is that of an evaporating black hole, which is well 
 described by the advanced/ingoing Vaidya metric near the 
 horizon~\cite{Bardeen:1981zz}.  In this picture, the change 
 of the local gravitational field supplies the energy of the 
 outgoing Hawking quanta created around $r \sim r_{\rm z}$.}

\begin{figure}[t]
\centering
  \subfigure{\includegraphics[clip,width=.49\textwidth]{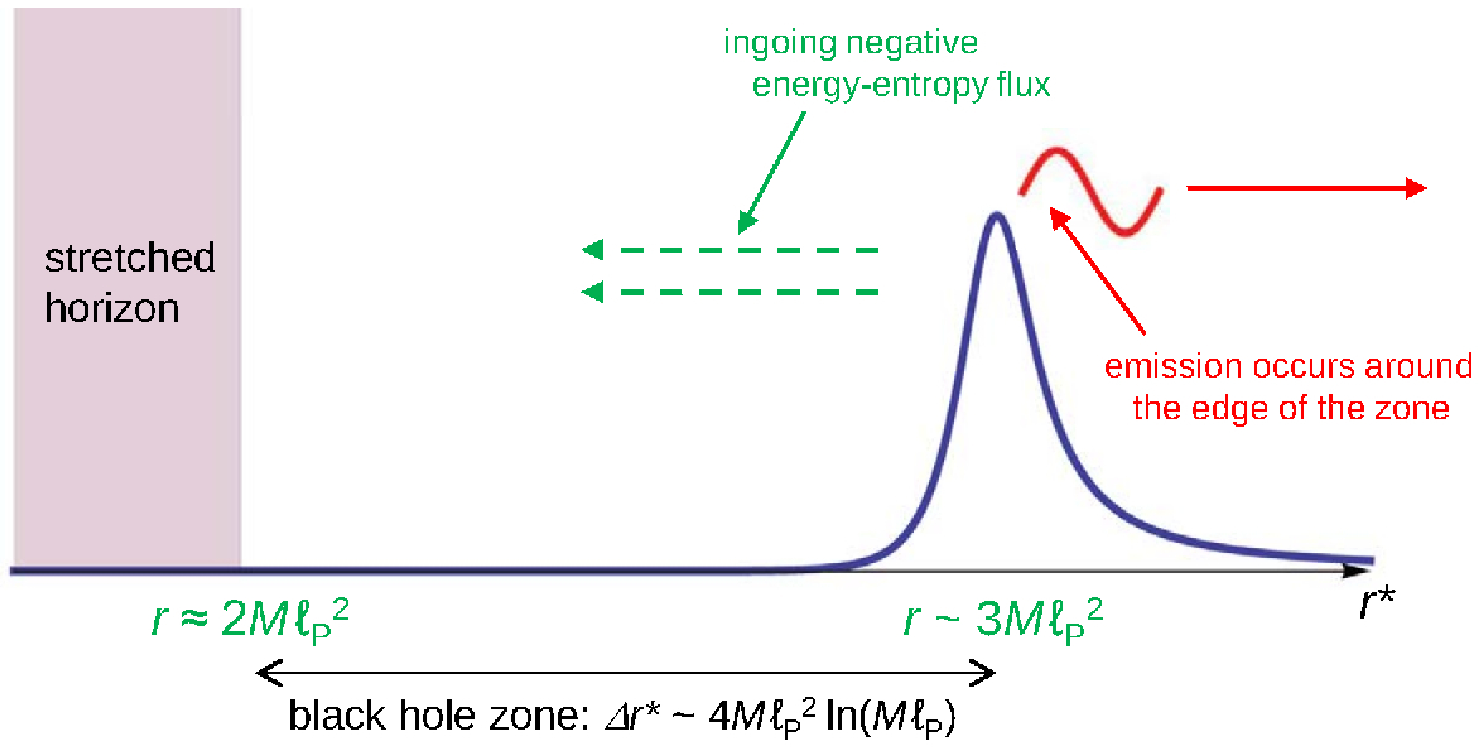}}
  \subfigure{\includegraphics[clip,width=.49\textwidth]{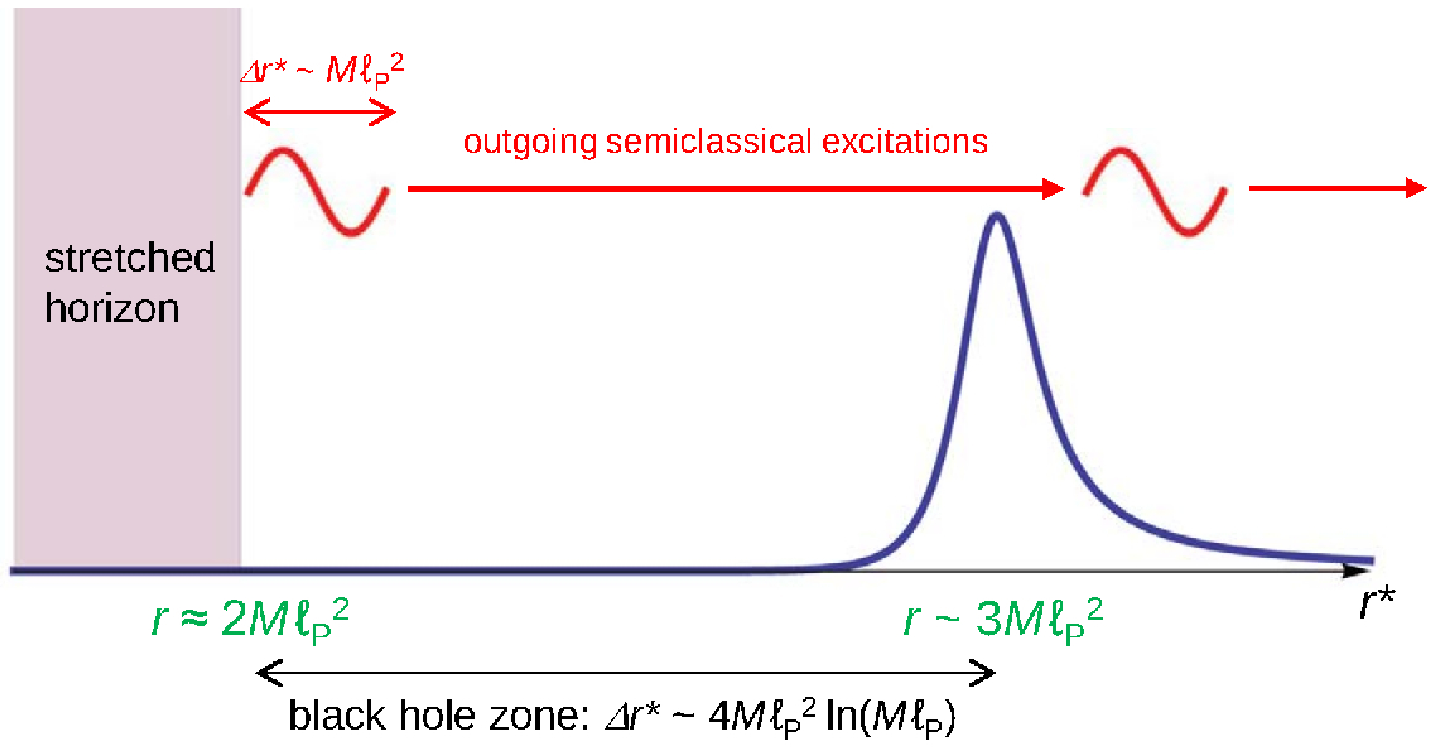}}
\caption{The information transfer from an evaporating black hole 
 occurs through negative energy-entropy excitations, created as 
 a backreaction of Hawking emission occurring around the edge of 
 the zone (left). This can be contrasted with the picture in which 
 outgoing positive energy-entropy excitations carry information 
 from the stretched horizon to the far region (right).}
\label{fig:info-trans}
\end{figure}

With the microscopic emission process in Eq.~(\ref{eq:toy-3}), 
a generic black hole state evolves unitarily as described by 
Page~\cite{Page:1993wv}; in particular, the entanglement between 
the black hole and the emitted Hawking radiation follows the Page 
curve.  As we have seen, the transfer of information from a black 
hole occurs through a negative entropy flux in the zone, carried 
by ingoing negative energy excitations on a background that can 
be viewed as static over the timescale of $O(M l_{\rm P}^2)$. 
This picture is different from that envisioned in 
Refs.~\cite{Almheiri:2012rt,Marolf:2013dba,Almheiri:2013hfa}, 
which assumed that information is carried from the stretched 
horizon to the edge of the zone by outgoing modes described 
within semiclassical theory; see Fig.~\ref{fig:info-trans}.

It is important that the transfer of information to semiclassical 
modes occurs mostly around the edge of the zone, not throughout the 
zone region.  For $\Delta$ sufficiently larger than $T_{\rm H}$, 
this condition is guaranteed because production of semiclassical 
modes in the bulk of the zone is suppressed by
\begin{equation}
  \epsilon \sim \biggl( \frac{\Delta}{T_{\rm H}} \biggr)^2 
    \, e^{-\frac{\Delta}{T_{\rm H}}},
\label{eq:suppress}
\end{equation}
where the first factor appears because for higher energies larger 
angular momentum modes can escape the zone.  This might, however, 
raise the following question.  Since the separation energy, 
$\Delta$, between the hard and soft modes is somewhat arbitrary, 
what happens if we artificially lower $\Delta$ down to $\approx 
T_{\rm H}$?  In this case, ``hard modes'' defined with respect 
to the lowered $\Delta$ seem to be produced throughout the zone. 
However, most of the produced modes cannot propagate to the 
edge of the zone.  The ``mean free path'' of these modes is 
of $O(M l_{\rm P}^2)$ in $r^*$, so that only the modes produced 
around the edge of the zone will escape to ambient space 
without being reabsorbed by the bath of the soft modes.  This 
is a manifestation of the fact that the modes with frequency 
$\lesssim T_{\rm H}$ must be viewed as soft modes, implying that 
$\Delta$ should be taken sufficiently larger than $T_{\rm H}$.

We conclude that semiclassical Hawking quanta must be regarded 
as emitted at the edge of the zone, where information stored 
in the soft modes---spacetime---is transferred to outgoing 
semiclassical degrees of freedom---Hawking quanta.  In fact, 
it is natural for such special dynamics to occur in this particular 
region, since this is where the near-horizon, Rindler-like space 
is ``patched'' to the asymptotic, Minkowski-like space in the 
semiclassical picture.

In the timescale of $O(M l_{\rm P}^2)$, the energy of the 
emitted Hawking quantum can only be measured with precision 
of $O(1/M l_{\rm P}^2)$.  Does this mean that the standard 
calculation of the gray-body factor is untrustable?  It does 
not.  Since the rate of Hawking emission is very small, the change 
of the black hole mass is very slow.  For example, after a very 
long time of $O(M^2 l_{\rm P}^3) \gg M l_{\rm P}^2$, the fractional 
change of the black hole mass is only of $O(1/M l_{\rm P}) \ll 1$. 
We then expect that the black hole state keeps taking the form 
of Eq.~(\ref{eq:BH-states}) with slowly varying $M$ through this 
long period, and we can apply the standard calculation for this 
long time, obtaining the result with the error of energy only 
of $O(1/M^2 l_{\rm P}^3) \ll 1/M l_{\rm P}$.  We find that 
the standard calculation of the gray-body factor, such as 
that in Ref.~\cite{Page:1976df}, can be trusted with precision, 
parametrically, of $O(1/\tau_{\rm BH}) \approx O(1/M^3 l_{\rm P}^4)$ 
down to zero energy in the spectrum.  This situation is quite 
different from the modes inside the zone, where the modes with 
frequency $\lesssim 1/M l_{\rm P}^2$ interact with the thermal 
bath in the timescale of $O(M l_{\rm P}^2)$; hence these modes 
must be viewed as soft modes, which do not have a structure 
beyond thermality at the semiclassical level.

\subsubsection{Evolution of microscopic entanglement}
\label{subsubsec:micro}

The Hawking emission process at the microscopic level, 
Eq.~(\ref{eq:toy-3}), indicates that a Hawking quantum shortly 
after the emission is entangled with the soft mode degrees of 
freedom.  How does this entanglement evolve?

To address this question, we first argue that the dimension of 
the Hilbert space associated with the hard modes is exponentially 
smaller than that associated with the soft modes.  In particular, 
the coarse-grained---or thermal---entropy associated with the hard 
modes is given by either
\begin{equation}
  S_{\rm hard} \approx O(1) 
    \biggl( \frac{{\cal A}(M)}{l_{\rm P}^2} \biggr)^p;
\qquad
  p < 1,
\label{eq:S_hard-1}
\end{equation}
or
\begin{equation}
  S_{\rm hard} = c\, \frac{{\cal A}(M)}{l_{\rm P}^2};
\qquad
  c \ll \frac{1}{4}.
\label{eq:S_hard-2}
\end{equation}
Here and below, we suppress the argument $M$ from entropies.  In the 
language of quantum information in holography, this amounts to saying 
that the dimension of the code subspace~\cite{Almheiri:2014lwa} 
erected on a black hole background (and representing the region 
$r \geq r_{\rm s}$) is exponentially smaller than that of the 
Hilbert space associated with the background geometry.  We 
strongly suspect that this is indeed the case, as anticipated 
earlier~\cite{'tHooft:1993gx,Nomura:2013lia}, but we may instead 
take it as an assumption of the framework.  Note that in 
Refs.~\cite{Nomura:2014woa,Nomura:2014voa,Nomura:2016qum}, 
it was argued that different black hole microstates must be 
viewed as different ``microscopic geometries.''  This corresponds 
to the statement that soft modes cannot be represented as degrees 
of freedom specifying states within a code subspace in a way 
that subsystem recovery is possible; only the hard and far 
modes can be represented in such a manner (unless an effective 
description having the second boundary is adopted; see 
Section~\ref{subsec:interior}).  Below, we assume that 
Eq.~(\ref{eq:S_hard-1}) or (\ref{eq:S_hard-2}) is true.

Without any dynamics swapping entanglement in the far region, 
the purifiers of the Hawking quanta emitted earlier keep being 
the microstates of the black hole.  The state of the combined 
black hole and Hawking radiation system at a given time is then
\begin{equation}
  \ket{\Psi(M)} = \sum_E \sum_{i_E = 1}^{{\cal N}(M-E)} 
    \sum_{a = 1}^{e^{S_{\rm rad}}} 
    c_{E i_E a} \ket{E} \ket{\psi_{i_E}(M-E)} \ket{r_a};
\qquad
  \sum_E \sum_{i_E = 1}^{{\cal N}(M-E)} 
    \sum_{a = 1}^{e^{S_{\rm rad}}} |c_{E i_E a}|^2 = 1,
\label{eq:BH-rad}
\end{equation}
where $\ket{r_a}$ represent orthonormal states for the radiation, 
and $S_{\rm rad}$ is its thermal entropy.  The density matrix for 
the hard modes is given by
\begin{align}
  \rho_{\rm H}(M) &= \sum_E \left( \sum_{i_E = 1}^{{\cal N}(M-E)} 
    \sum_{a = 1}^{e^{S_{\rm rad}}}  |c_{E i_E a}|^2 \right) 
    \ket{E} \bra{E}
\nonumber\\
  &\simeq \frac{1}{\sum_E e^{-\frac{E}{T_{\rm H}}}} 
    \sum_E e^{-\frac{E}{T_{\rm H}}} \ket{E} \bra{E},
\label{eq:rho_hard}
\end{align}
where in the second line, we have used the fact that the size of 
$|c_{E i_E a}|^2$ is statistically given by%
\footnote{We are not concerned with the logarithmic correction 
 to the black hole entropy arising from the $\varDelta M/M$ factor 
 in Eq.~(\ref{eq:S_BH}), so we will identify ${\cal N}(M)$ as 
 $e^{S_{\rm BH}}$ and similarly for other numbers of microstates.}
\begin{equation}
  |c_{E i_E a}|^2 \sim \frac{1}{e^{S_{\rm BH} + S_{\rm rad}} 
    \sum_E e^{-\frac{E}{T_{\rm H}}}}.
\label{eq:c_Eia}
\end{equation}
This takes the same form as Eq.~(\ref{eq:BH-thermal}), so that 
the physics of the hard modes is still described by standard 
semiclassical theory.  The density matrix for the emitted Hawking 
radiation, at the time the black hole has mass $M$, is
\begin{equation}
  \rho_{\rm R}(M) = \sum_{a,b = 1}^{e^{S_{\rm rad}}} 
    \left( \sum_E \sum_{i_E = 1}^{{\cal N}(M-E)} 
    c_{E i_E a} c^*_{E i_E b} \right) \ket{r_a} \bra{r_b}.
\label{eq:rho_rad}
\end{equation}
For a quantum chaotic dynamics of the black hole, the von~Neumann 
entropy of this density matrix follows the Page curve as the black 
hole evaporation progresses.

It is instructive to study the structure of tripartite 
entanglement in Eq.~(\ref{eq:BH-rad}) further.  This expression 
tells us that the state of the hard modes, $\rho_{\rm H}(M)$ 
in Eq.~(\ref{eq:rho_hard}), is purified by the states of the 
combined system of soft modes and radiation
\begin{equation}
  \ket{\tilde{E}} 
  = \frac{1}{\sqrt{z_E}} \sum_{i_E = 1}^{{\cal N}(M-E)} 
    \sum_{a = 1}^{e^{S_{\rm rad}}} 
    c_{E i_E a} \ket{\psi_{i_E}(M-E)} \ket{r_a};
\qquad
  z_E = \sum_{i_E = 1}^{{\cal N}(M-E)} 
    \sum_{a = 1}^{e^{S_{\rm rad}}} |c_{E i_E a}|^2.
\label{eq:mirror-st}
\end{equation}
Suppose that the coarse-grained/thermal entropies of the three 
sectors satisfy
\begin{equation}
  S_{\rm hard} \ll S_{\rm soft} \approx S_{\rm BH}, S_{\rm rad},
\label{eq:S-rel}
\end{equation}
which is expected to be valid throughout the history of black hole 
evolution, except possibly in the earliest time when $S_{\rm rad} 
\lesssim S_{\rm hard}$ (if the black hole equilibrates before 
$S_{\rm rad}$ becomes larger than $S_{\rm hard}$).  Let us now 
perform the Schmidt decomposition in the space given by the soft 
mode and radiation states {\it for each $E$}:
\begin{equation}
  \ket{\tilde{E}} = \sum_{i_E=1}^{{\cal N}_E} 
    \gamma_{E i_E} \ket{\psi_{i_E}} \ket{r_{i_E}};
\qquad
  \sum_{i_E=1}^{{\cal N}_E} \gamma_{E i_E}^2 = 1,
\label{eq:Schmidt}
\end{equation}
where
\begin{equation}
  {\cal N}_E = {\rm min}\{ {\cal N}(M-E), e^{S_{\rm rad}} \},
\label{eq:N}
\end{equation}
and $\ket{\psi_{i_E}}$ without the argument $M-E$ represent Schmidt 
basis states.  In the above expression, we have kept each entropy 
only at the leading relevant order in expansion in inverse powers 
of $M l_{\rm P}$.  By construction, the states of the soft modes 
as well as those of the radiation in this basis are orthonormal 
for each $E$:
\begin{equation}
  \inner{\psi_{i_E}}{\psi_{j_E}} = \delta_{i_E j_E},
\qquad
  \inner{r_{i_E}}{r_{j_E}} = \delta_{i_E j_E},
\label{eq:Sch-orthonorm}
\end{equation}
and all the coefficients in Eq.~(\ref{eq:N}) are real and 
non-negative, $\gamma_{E i_E} \geq 0$.

In general, soft mode states corresponding to different $E$ are 
orthogonal
\begin{equation}
  \inner{\psi_{i_E}}{\psi_{j_{E'}}} = 0 
\quad
  \mbox{for } E \neq E',
\label{eq:psi-ortho}
\end{equation}
but the same is not necessarily true for radiation states.  However, 
the inner product of two generic radiation states is suppressed 
by the large dimension of the radiation Hilbert space:
\begin{equation}
  |\inner{r_{i_E}}{r_{j_{E'}}}| 
  \approx O\biggl(\frac{1}{e^{\frac{1}{2}S_{\rm rad}}}\biggr) \ll 1
\quad
  \mbox{for } E \neq E'.
\label{eq:r-ortho}
\end{equation}
Here, motivated by the expectation that the dynamics of the black 
hole is quantum chaotic, we have assumed that the distributions 
of ${\cal N}_E$ states $\ket{r_{i_E}}$ and ${\cal N}_{E'}$ states 
$\ket{r_{j_{E'}}}$ are uncorrelated in the radiation Hilbert space 
of dimension $e^{S_{\rm rad}}$.  This allows us to view that the 
state of the entire system takes the form
\begin{equation}
  \ket{\Psi(M)} = \sum_E \sqrt{z_E} 
    \sum_{i_E=1}^{{\cal N}_E} \gamma_{E i_E} 
    \ket{E} \ket{\psi_{i_E}} \ket{r_{i_E}},
\label{eq:GHZ}
\end{equation}
with all the $\ket{\psi_{i_E}}$'s as well as all the 
$\ket{r_{i_E}}$'s being (approximately) orthogonal.  Here, $z_E$ 
is defined in Eq.~(\ref{eq:mirror-st}).

The entanglement structure in Eq.~(\ref{eq:GHZ}) is reminiscent of 
the GHZ form~\cite{GHZ}.  To see the significance of this statement, 
let us trace out the radiation degrees of freedom and obtain the 
reduced density matrix describing the hard and soft modes
\begin{align}
  \rho_{\rm HS}(M) &= {\rm Tr}_{\rm rad} \ket{\Psi(M)} \bra{\Psi(M)}
\nonumber\\
  &= \sum_E \sum_{i_E=1}^{{\cal N}_E} z_E \gamma_{E i_E}^2 
    \ket{E} \ket{\psi_{i_E}} \bra{E} \bra{\psi_{i_E}} 
\nonumber\\
  & {} \quad + \sum_{\substack{E,E' \\ E \neq E'}} 
    \sum_{i_E=1}^{{\cal N}_E} \sum_{i'_{E'}=1}^{{\cal N}_{E'}} 
    \sqrt{z_E z_{E'}}\, \gamma_{E i_E} \gamma_{E' i'_{E'}} 
    O\left( \frac{1}{e^{\frac{1}{2}S_{\rm rad}}} \right)
    \ket{E} \ket{\psi_{i_E}} \bra{E'} \bra{\psi_{i'_{E'}}},
\label{eq:rho_HS}
\end{align}
where we have assumed generic sizes for the coefficients
\begin{equation}
  |c_{E i_E a}|^2 \sim \frac{1}{e^{S_{\rm BH} + S_{\rm rad}} 
    \sum_E e^{-\frac{E}{T_{\rm H}}}},
\qquad
  \gamma_{E i_E}^2 \sim \frac{1}{{\cal N}_E},
\qquad
  z_E \sim 
    \frac{e^{-\frac{E}{T_{\rm H}}}}{\sum_E e^{-\frac{E}{T_{\rm H}}}},
\label{eq:generic}
\end{equation}
and ignored the irrelevant factor of $e^{S_{\rm hard}}$.  Note that 
the phases of the second term in Eq.~(\ref{eq:rho_HS}) are random 
because of random phases from inner products between different 
radiation states.  Similarly, 
we can trace out the soft modes and obtain the reduced density 
matrix for the hard modes and the radiation
\begin{align}
  \rho_{\rm HR}(M) &= {\rm Tr}_{\rm soft} \ket{\Psi(M)} \bra{\Psi(M)}
\nonumber\\
  &= \sum_E \sum_{i_E=1}^{{\cal N}_E} z_E \gamma_{E i_E}^2 
    \ket{E} \ket{r_{i_E}} \bra{E} \bra{r_{i_E}}.
\label{eq:rho_HR}
\end{align}
This takes the diagonal form.

The expressions in Eq.~(\ref{eq:rho_HS},~\ref{eq:rho_HR}) indicate 
that the correlation of the hard modes with either of the soft modes 
or radiation is (essentially) classical.  For Eq.~(\ref{eq:rho_HR}) 
this is obvious, and for Eq.~(\ref{eq:rho_HS}) it is due to the 
extra $e^{-S_{\rm rad}/2}$ factor in the second term originating 
from Eq.~(\ref{eq:r-ortho}).  It is striking that the hard modes 
can be purified only when we consider the combined system of the 
soft modes and the early radiation.  This is the case regardless 
of the relative size between $S_{\rm soft}$ and $S_{\rm rad}$, 
i.e.\ whether the age of the black hole is younger or older than 
the Page time.

Incidentally, the correlation between the soft modes and the radiation 
given by the reduced density matrix
\begin{equation}
  \rho_{\rm SR}(M) = \sum_E z_E \sum_{i_E,i'_E=1}^{{\cal N}_E} 
    \gamma_{E i_E} \gamma_{E i'_E} 
    \ket{\psi_{i_E}} \ket{r_{i_E}} \bra{\psi_{i'_E}} \bra{r_{i'_E}},
\label{eq:rho_SR}
\end{equation}
is generally quantum mechanical as required for the unitary 
evolution of the black hole.  This is a feature that makes the 
entanglement structure of Eq.~(\ref{eq:GHZ}) different from the 
true GHZ form.

\section{Interior Spacetime}
\label{sec:interior}

In this section, we study what happens to an object falling into 
an evaporating black hole.  We analyze how the interior spacetime 
manifests itself in the microscopic description of the black hole. 
We also discuss relations of this picture with the resolution of 
the cloning paradox.

\subsection{An object falling into a black hole}
\label{subsec:falling}

\begin{figure}[t]
\begin{center}
\centering
  \subfigure{\includegraphics[clip,width=.49\textwidth]{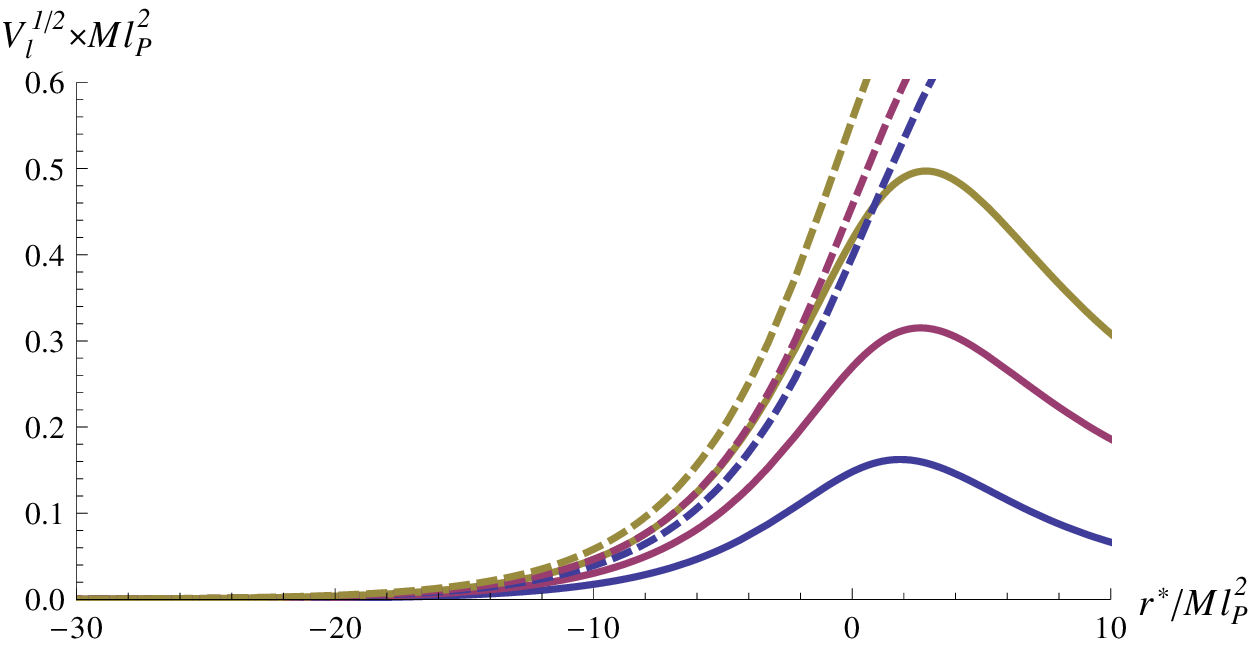}}
  \subfigure{\includegraphics[clip,width=.49\textwidth]{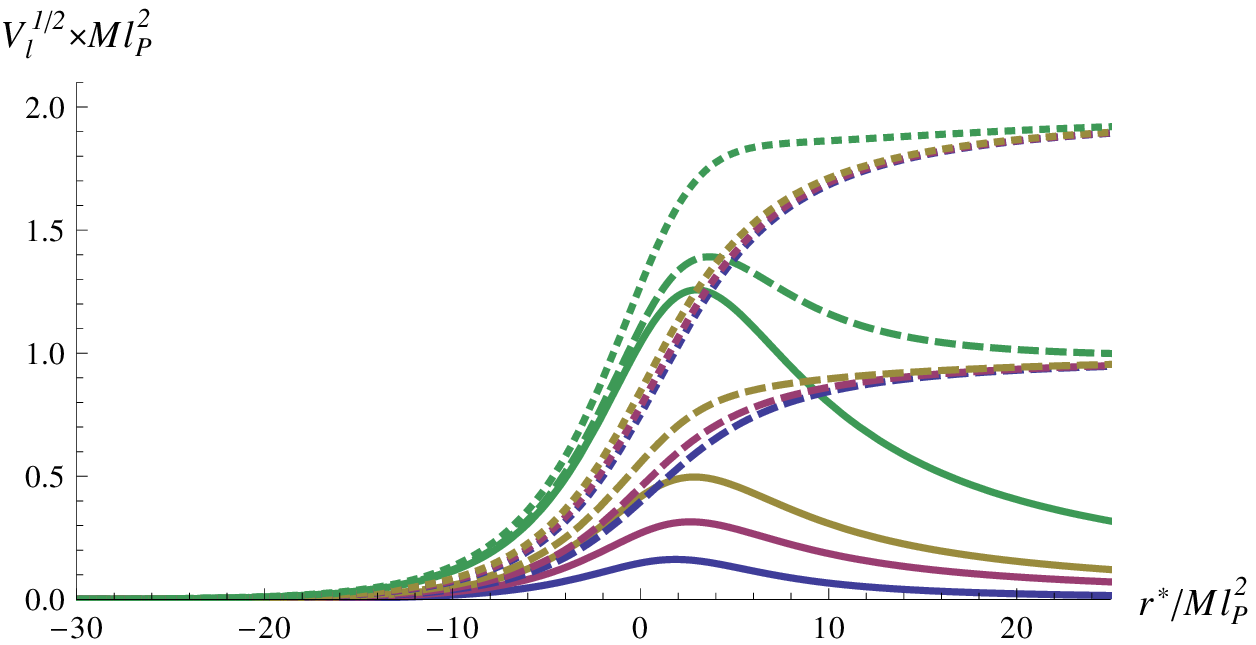}}
\end{center}
\caption{The potential $\sqrt{V_l(r^*)}$ in units of $1/Ml_{\rm P}^2$, 
 plotted as a function of $r_*/Ml_{\rm P}^2$.  In the left panel, 
 solid and dashed lines represent the potential for $\mu = 0$ and 
 $1/Ml_{\rm P}^2$, respectively; for each value of $\mu$, $l = 0,1,2$ 
 are plotted (from bottom to top).  In the right panel, the cases 
 of $\mu Ml_{\rm P}^2 = 0, 1, 2$ are plotted on a different scale 
 (solid, dashed, and dotted, respectively), now for $l = 0,1,2,5$ 
 for each value of $\mu$ (from bottom to top).}
\label{fig:potential}
\end{figure}

We first analyze an object falling into a black hole.  Consider a 
scalar field $\varphi$ of mass $\mu$.  In the tortoise coordinates, 
its action can be written as
\begin{align}
  I =& \frac{1}{2} \int\!dt\, dr^* d\theta\, d\phi\, 
    \biggl( 1 - \frac{2 M l_{\rm P}^2}{r} \biggr)\, r^2 \sin\theta 
\nonumber\\
  &\quad \times \left[ \frac{1}{1 - \frac{2 M l_{\rm P}^2}{r}} 
      \biggl\{ \biggl( \frac{\partial \varphi}{\partial t} \biggr)^2 
      - \biggl( \frac{\partial \varphi}{\partial r^*} \biggr)^2 
        \biggr\} - \frac{1}{r^2} 
      \biggl( \frac{\partial \varphi}{\partial \theta} \biggr)^2 
      - \frac{1}{r^2 \sin^2\!\theta} 
      \biggl( \frac{\partial \varphi}{\partial \phi} \biggr)^2 
      - \mu^2 \varphi^2 \right],
\label{eq:I_phi}
\end{align}
where $\theta$ and $\phi$ are angular coordinates, and $r$ is 
a function of $r^*$ determined by Eq.~(\ref{eq:tortoise}).  By 
rescaling the field and decomposing into spherical harmonics
\begin{equation}
  \varphi(t,r^*,\theta,\phi) 
  = \frac{1}{r} \sum_{l,m} \chi_{lm}(t,r^*) Y_{lm}(\theta,\phi),
\label{eq:decomp}
\end{equation}
we obtain
\begin{equation}
  I = \frac{1}{2} \sum_{l,m} \int\!dt\, dr^* 
    \left[ \biggl( \frac{\partial \chi_{lm} }{\partial t} \biggr)^2 
      - \biggl( \frac{\partial \chi_{lm}}{\partial r^*} \biggr)^2 
        - V_l(r^*)\, \chi_{lm}^2 \right],
\label{eq:I_chi}
\end{equation}
where
\begin{equation}
  V_l(r^*) = \biggl( 1 - \frac{2 M l_{\rm P}^2}{r} \biggr) 
    \biggl( \frac{l(l+1)}{r^2} + \frac{2 M l_{\rm P}^2}{r^3} 
    + \mu^2 \biggr).
\label{eq:V_l}
\end{equation}
The equation of motion for a mode of frequency $\omega$ is then
\begin{equation}
  -\frac{\partial^2 \chi_{lm}}{(\partial r^*)^2} + 
    V_l(r^*)\, \chi_{lm} = \omega^2 \chi_{lm}.
\label{eq:eom}
\end{equation}
Note that $\omega$ is a conserved quantity, since the action is 
invariant under translation in $t$.  For convenience, we plot 
$\sqrt{V_l(r^*)}$ in Fig.~\ref{fig:potential} for some small 
values of $l$ and $\mu$.

Let us first consider the case with $\mu = 0$.  Suppose we drop 
a wavepacket into a black hole from the outside of the zone, with 
$r_{\rm init} \approx O(M l_{\rm P}^2) > r_{\rm z}$.  We assume 
that the characteristic width $d$ of the wavepacket in the angular 
directions, which we call the transverse directions, is much 
smaller than the radius of the black hole:\ $d \ll R$, where 
$R = 2 M l_{\rm P}^2$.  In this case, uncertainty in the transverse 
momentum is of order $\varDelta p_\perp \approx 1/d$, and the 
frequencies of the modes composing the wavepacket have a spread
\begin{equation}
  \varDelta \omega \approx \frac{1}{d} \gg \frac{1}{M l_{\rm P}^2}.
\label{eq:del-omega}
\end{equation}
This implies that the spread of the energy is much larger than the 
separation energy between the hard and soft modes
\begin{equation}
  \varDelta \omega \approx O\biggl(\frac{R}{d}\biggr) \Delta 
  \gg \Delta.
\label{eq:del-omega-2}
\end{equation}
In terms of angular momentum, the spread is $\varDelta L \approx 
O(R \varDelta p_\perp)$, giving
\begin{equation}
  \varDelta l \approx O\biggl(\frac{R}{d}\biggr) \gg 1.
\label{eq:del-l}
\end{equation}
Note that while the peak of the potential $V_l(r^*)$, located around 
$|r^*| \lesssim O(M l_{\rm P}^2)$, is higher for larger $l$, given 
by $V_{l,{\rm max}} \simeq (1/27) (l^2/M^2 l_{\rm P}^4)$, the 
contributions to the energy in Eq.~(\ref{eq:del-omega-2}) allow 
for the wavepacket to enter into the zone over the potential 
barrier.

The situation for $\mu \neq 0$ is similar.  In this case, 
frequencies receive the contribution from the rest mass:
\begin{equation}
  \omega^2 = \mu^2 + p^2.
\label{eq:omega-mass}
\end{equation}
For $\mu \gtrsim (1/\sqrt{12})l/M l_{\rm P}^2$, the effect of 
gravitation acting on the rest mass makes the barrier disappear; 
otherwise, the height of the bump is as before, $\varDelta V_l 
\simeq (1/27) (l^2/M^2 l_{\rm P}^4)$.  The analysis in the case 
of $\mu = 0$ applies essentially with $\omega$ replaced by $p$. 
In particular, the spreads of various quantities are
\begin{equation}
  \sqrt{\varDelta \omega^2} \approx \sqrt{\varDelta p^2} 
  \approx O\biggl(\frac{R}{d}\biggr) \Delta \gg \Delta.
\label{eq:spreads}
\end{equation}
Note that it is the square of the frequency that is relevant for the 
dynamics; see Eq.~(\ref{eq:eom}).  In fact, the term involving the 
mass in $V_l(r^*)$ is negligible compared with $T_{\rm H}^2$ in the 
region $T_{\rm loc}(r) \gg \mu$.  Therefore, any elementary particle, 
for which $\mu < l_{\rm s}$, can be regarded as massless near the 
stretched horizon.

As we discussed in Section~\ref{sec:distant}, modes 
with frequencies smaller than $\Delta$ cannot be 
discriminated at the semiclassical level.  The relations 
in Eqs.~(\ref{eq:del-omega-2},~\ref{eq:spreads}), however, 
show that the details of a particle, such as the form of the 
wavefunction, can still be described at this level as long as 
the particle is localized within the lengthscale $d \ll R$ in 
the transverse directions.  The error coming from neglecting the 
soft mode contribution to the wavefunction is suppressed by $d/R$, 
consistent with the equivalence principle.  A similar statement 
also applies to an object consisting of many particles, whose 
size is determined not by the spread of the wavefunction but by 
interactions between the constituents:\ as long as its size in 
the transverse directions is sufficiently smaller than $R$, its 
dynamics can be well described at the semiclassical level.

This implies that to describe the dynamics of such ``small'' 
objects, it is sufficient to consider operators acting on the 
``$E$ space'' in Eq.~(\ref{eq:BH-rad}):
\begin{equation}
  {\cal O}\, \ket{E} \ket{\psi_{i_E}(M-E)} \ket{r_a} 
  = \sum_{E'} o_{E E'} \ket{E'} \ket{\psi_{i_E}(M-E)} \ket{r_a},
\label{eq:semicl-op}
\end{equation}
where $o_{E E'}$ are the elements of a matrix defined in the space 
spanned by $\{ \ket{E} \}$.  In particular, small objects in the 
zone, $r_{\rm s} \leq r \leq r_{\rm z}$, can be described by the 
thermal density matrix in Eq.~(\ref{eq:rho_hard}) and operators 
acting on it through
\begin{equation}
  \ket{E} \rightarrow \sum_{E'} o_{E E'} \ket{E'},
\label{eq:zone}
\end{equation}
without referring to the state of the soft modes or the early 
radiation.  In fact, the change of the density matrix of the hard 
modes due to Hawking emission, which occurs through interactions 
with the soft modes, is sufficiently slow that the description 
based on these modes can be used for a timescale much longer 
than $O(M l_{\rm P}^2)$ (where possible changes occurring through 
a long time period can be treated adiabatically).  It is this 
description that we call semiclassical theory.

\subsection{Emergence of the interior}
\label{subsec:interior}

What happens when a falling object reaches the stretched horizon? 
In the viewpoint of a distant observer, the information about the 
object will be transferred to the excitations of the stretched 
horizon, which will eventually be resolved into states of the 
soft modes.  However, there is another, coarse-grained description 
applicable only to certain coarse-grained degrees of freedom in 
a certain limited regime.  This leads to the emergence of spacetime 
inside the horizon.  In fact, this is the only sense in which the 
concept of the black hole interior can come out from the microscopic 
point of view.  In this subsection, we study this issue.

\subsubsection{Two-sided description}
\label{subsubsec:two-sided}

Consider a black hole of mass $M$.  We label the states for 
the hard modes, $\ket{E}$, in terms of the occupation numbers, 
$n_\alpha$, for each mode $\alpha$:
\begin{equation}
  \ket{E} \rightarrow \ket{\{ n_\alpha \}}.
\label{eq:n_xi}
\end{equation}
Here, $\alpha$ collectively denotes the species, frequency, 
and angular-momentum quantum numbers of the mode.  The state 
of the combined system of the black hole and radiation, 
Eq.~(\ref{eq:BH-rad}), can then be written as
\begin{equation}
  \ket{\Psi(M)} = \sum_n \sum_{i_n = 1}^{{\cal N}(M-E_n)} 
    \sum_a c_{n i_n a} \ket{\{ n_\alpha \}} 
    \ket{\psi_{i_n}(M-E_n)} \ket{r_a};
\qquad
  \sum_n \sum_{i_n = 1}^{{\cal N}(M-E_n)} \sum_a |c_{n i_n a}|^2 = 1,
\label{eq:BH-occup}
\end{equation}
where $n \equiv \{ n_\alpha \}$ represents the set of all 
occupation numbers, and $E_n$ is the energy of the state 
$\ket{\{ n_\alpha \}}$ as measured in the asymptotic region 
within precision $\Delta$.  Important operators of the form of 
Eq.~(\ref{eq:semicl-op}) are annihilation and creation operators
\begin{align}
  b_\gamma &= \sqrt{n_\gamma}\, 
    \ket{\{ n_\alpha - \delta_{\alpha\gamma} \}} \bra{\{ n_\alpha \}} 
    \otimes {\bf 1} \otimes {\bf 1},
\label{eq:ann}\\
  b_\gamma^\dagger &= \sqrt{n_\gamma + 1}\, 
    \ket{\{ n_\alpha + \delta_{\alpha\gamma} \}} \bra{\{ n_\alpha \}} 
    \otimes {\bf 1} \otimes {\bf 1},
\label{eq:cre}
\end{align}
where ${\bf 1}$ represents the fact that the operators do not act 
on soft mode or radiation states.  These comprise operators in 
the semiclassical theory describing physics in the zone region, 
$r_{\rm s} \leq r \leq r_{\rm z}$.  As we have seen in 
Section~\ref{subsec:falling}, a small object falling toward 
the horizon can be well described by the configuration of the 
hard modes, i.e.\ these operators acting on the black hole state.

As we will see explicitly in Section~\ref{subsubsec:inside}, we 
can describe what happens to such a small, falling object without 
knowing the detailed states of the soft modes or early radiation. 
We can therefore coarse-grain these states as
\begin{equation}
  \sum_{i_n = 1}^{{\cal N}(M-E_n)} \sum_a c_{n i_n a} 
    \ket{\psi_{i_n}(M-E_n)} \ket{r_a} 
  \quad\longrightarrow\quad
  \sqrt{\sum_{i_n = 1}^{{\cal N}(M-E_n)} \sum_a\, 
    |c_{n i_n a}|^2}\, \ketc{\{ n_\alpha \}}.
\label{eq:coarse}
\end{equation}
Here, we have used the same label as the hard mode state to specify 
the coarse-grained state, which we denote by the double ket symbol, 
and the coefficient in the right-hand side arises because we have 
taken $\ketc{\{ n_\alpha \}}$ to be a normalized state.  At the 
coarse-grained level, the detailed structures of $c_{n i_n a}$'s 
are not important, so this coefficient can be written as
\begin{equation}
  \sqrt{\sum_{i_n = 1}^{{\cal N}(M-E_n)} \sum_a\, 
    \langle |c_{n i_n a}|^2 \rangle} 
  \approx \sqrt{\frac{{\cal N}(M-E_n)}{\sum_n {\cal N}(M-E_n)}}
  = \frac{e^{-4\pi M E_n l_{\rm P}^2}}
      {\sqrt{\sum_n e^{-8\pi M E_n l_{\rm P}^2}}} 
  = \frac{e^{-\frac{E_n}{2 T_{\rm H}}}}
      {\sqrt{\sum_n e^{-\frac{E_n}{T_{\rm H}}}}},
\label{eq:c-coeff}
\end{equation}
where
\begin{equation}
    \langle |c_{n i_n a}|^2 \rangle 
  \approx \frac{1}{\sum_n \sum_a {\cal N}(M-E_n)},
\label{eq:c-exp}
\end{equation}
is the characteristic size of $|c_{n i_n a}|^2$, obtained from the 
normalization condition in Eq.~(\ref{eq:BH-occup}).  The state in 
Eq.~(\ref{eq:BH-occup}) can then be written as
\begin{equation}
  \ketc{\Psi(M)} 
  = \frac{1}{\sqrt{\sum_n e^{-\frac{E_n}{T_{\rm H}}}}} 
    \sum_n e^{-\frac{E_n}{2 T_{\rm H}}} 
    \ket{\{ n_\alpha \}} \ketc{\{ n_\alpha \}},
\label{eq:BH-coarse}
\end{equation}
which takes the form of the standard thermofield double state in 
the two-sided black hole picture~\cite{Unruh:1976db,Israel:1976ur}, 
although $\ket{\{ n_\alpha \}}$ here represent the states only of 
the hard modes.%
\footnote{In this picture, negative energy-entropy excitations of 
 Section~\ref{subsec:Hawking} arising as backreaction of Hawking 
 emission appear as ``particles'' whose wavelengths are of the 
 order of the black hole horizon radius, or an ambiguity in 
 choosing a vacuum at this scale resulting from spacetime curvature. 
 We will ignore this effect, which is not important in discussing 
 physics of a small infalling object.}

We can now define the ``mirror operators'' acting on the 
coarse-grained states~\cite{Papadodimas:2012aq,Papadodimas:2013jku,%
Papadodimas:2015jra}:
\begin{align}
  \tilde{b}_\gamma &= {\bf 1} \otimes \sqrt{n_\gamma}\, 
    \ketc{\{ n_\alpha - \delta_{\alpha\gamma} \}} 
    \brac{\{ n_\alpha \}},
\label{eq:ann-m}\\
  \tilde{b}_\gamma^\dagger &= {\bf 1} \otimes \sqrt{n_\gamma + 1}\, 
    \ketc{\{ n_\alpha + \delta_{\alpha\gamma} \}} 
    \brac{\{ n_\alpha \}},
\label{eq:cre-m}
\end{align}
where the two factors represent operators acting on 
$\ket{\{ n_\alpha \}}$ (trivially) and $\ketc{\{ n_\alpha \}}$ 
in Eq.~(\ref{eq:BH-coarse}).  Note that, as emphasized 
in Refs.~\cite{Papadodimas:2012aq,Papadodimas:2013jku,%
Papadodimas:2015jra}, these operators cannot be defined in 
a state-independent manner at the microscopic level---if we 
want to define operators corresponding to Eqs.~(\ref{eq:ann-m},%
~\ref{eq:cre-m}) at the microscopic level, then we must do so in 
a way dependent on the microstate of a black hole.  The analysis 
in Section~\ref{subsec:Hawking} indicates that such state-dependent 
operators must act on {\it both} soft modes {\it and} early 
radiation, {\it regardless of the age of the black hole}.

In describing a small object falling inside the horizon, however, 
we need not explicitly consider microscopic operators.  This is 
because as long as we restrict our attention to a certain spacetime 
region, the dynamics dictating the infalling object can be described 
unitarily in the Fock space given by (not too many) operators in 
Eqs.~(\ref{eq:ann},~\ref{eq:cre},~\ref{eq:ann-m},~\ref{eq:cre-m}) 
acting on the state in Eq.~(\ref{eq:BH-coarse}).  In 
Section~\ref{subsubsec:inside} we will specify what 
the certain spacetime region means.

If Hawking radiation emitted earlier interacts with environment, it 
may transfer a part (or all) of its entanglement with the black hole 
to the environment.  The states $\ket{r_a}$ in Eq.~(\ref{eq:BH-occup}) 
must then include the environment as well.  In fact, $\ket{r_a}$ can 
be viewed in general as full states representing the region outside 
the zone.

What happens if a detector collects a large number of Hawking quanta 
and then enters into the black hole?  Imagine that early Hawking 
radiation interacts with a detector, leading to different pointer 
states $\ket{d_I}$.  By separating these states from $\ket{r_a}$, 
the state in Eq.~(\ref{eq:BH-occup}) can be written as
\begin{equation}
  \ket{\Psi(M)} = \sum_n \sum_{i_n = 1}^{{\cal N}(M-E_n)} 
    \sum_I \sum_{a_I} c_{n i_n I a_I} \ket{\{ n_\alpha \}} 
    \ket{\psi_{i_n}(M-E_n)} \ket{r_{a_I}} \ket{d_I},
\label{eq:BH-occup-2}
\end{equation}
where $\sum_n \sum_{i_n = 1}^{{\cal N}(M-E_n)} \sum_I \sum_{a_I} 
|c_{n i_n I a_I}|^2 = 1$.  To discuss what the detector finding 
a particular outcome $I$ will experience later, we may focus on 
the particular branch of the wavefunction
\begin{equation}
  \ket{\Psi_I(M)} = \frac{1}{\sqrt{z_I}} \sum_n 
    \sum_{i_n = 1}^{{\cal N}(M-E_n)} \sum_{a_I} c_{n i_n I a_I} 
    \ket{\{n_\alpha\}} \ket{\psi_{i_n}(M-E_n)} \ket{r_{a_I}} \ket{d_I},
\label{eq:Psi_I}
\end{equation}
where $z_I = \sum_n \sum_{i_n = 1}^{{\cal N}(M-E_n)} \sum_{a_I} 
|c_{n i_n I a_I}|^2$.  Generically, this does not affect the physics 
of the black hole, since the structure of Eq.~(\ref{eq:Psi_I}) 
is the same as that of Eq.~(\ref{eq:BH-occup}).  However, if 
the detector is carefully set up, it may be fully correlated 
with a particular configuration $\{ n'_\alpha \}$ of the hard 
modes after the measurement:\ $c_{n i_n I a_I} \approx 0$ for 
$n \neq \{ n'_\alpha \}$.  This seems to mean that when the detector 
enters the horizon, it would hit a ``firewall'' because the hard 
modes lack entanglement.  This, however, need not be the case.

Since the detector can interact with Hawking quanta only outside the 
zone, it takes time of order $4M l_{\rm P}^2 \ln(M l_{\rm P})$ for 
it to reach the stretched horizon.  (More generally, it takes time 
of order $4M l_{\rm P}^2 \ln(M l_{\rm P})$ for the outcome of the 
measurement to be communicated to the region near the stretched 
horizon.)  Therefore, if the equilibrium timescale between the hard 
and soft modes is of order
\begin{equation}
  t_{\rm eq} = 4Ml_{\rm P}^2 \ln(Ml_{\rm P}),
\label{eq:t_eq}
\end{equation}
or shorter, then the state of the system (without the detector 
included) takes the form of Eq.~(\ref{eq:BH-occup}) with generic 
$c_{n i_n a} = c_{n i_n I a_I}$ when the detector reaches the 
stretched horizon.  This implies that the detector sees a smooth 
horizon when it falls into the black hole.

The above analysis suggests that an operation acting only on 
early Hawking radiation---however complicated---cannot destroy 
the smoothness of the horizon.  This is a consequence of the 
entanglement structure described in Section~\ref{subsec:Hawking}.

\subsubsection{Effective theories of the interior}
\label{subsubsec:inside}

Suppose that the state of the entire system at a given Schwarzschild 
time $t = t_*$ has a black hole of mass $M$ with a small object in 
the zone falling toward the horizon.  To see what happens to this 
object, we may adopt the coarse-grained description given above, 
in which the state is given by
\begin{equation}
  \ketc{\Psi_0} \propto \prod_{i=1}^N
    \left( \sum_\gamma f_{i,\gamma} b_\gamma^\dagger \right) 
    \ketc{\Psi(M)},
\label{eq:Psi_0}
\end{equation}
where we have assumed that the object consists of $N$ particles, 
and $\gamma$ collectively denotes frequency as well as other 
discrete labels such as those for particle species and angular 
momenta.  The coefficients $f_{i,\gamma}$ are the weights needed 
to produce particle $i$ by superposing the creation operators 
$b_\gamma^\dagger$, and the coarse-grained black hole vacuum state 
$\ketc{\Psi(M)}$ is given by Eq.~(\ref{eq:BH-coarse}).  Note that 
we can always discriminate constituents of a semiclassical object 
from the thermal atmosphere, since they modulate the thermal density 
matrix of Eq.~(\ref{eq:rho_hard}) with energies larger than the 
energy spread of the black hole vacuum, $\varDelta E \approx 
1/Ml_{\rm P}^2$.

The question is:\ what does this object experience when it enters 
the horizon?  To answer this question, evolving the state in 
Schwarzschild time $t$ is of no use.  In such a description, the 
object hits the stretched horizon and is converted into excitations 
on the stretched horizon, which is then resolved into soft modes due 
to intrinsically stringy dynamics.  In the context of holography, 
this implies that the boundary time evolution cannot be used 
to provide the answer, since it corresponds to evolution in 
Schwarzschild time~\cite{Nomura:2018kji}.  To address the question, 
we need to ``evolve'' the state in a way related to the proper 
time seen by the object.

Since the coarse-grained black hole vacuum state $\ketc{\Psi(M)}$ 
takes the standard thermofield double form, the Fock space built on 
it can be viewed as representing excitations on a background of the 
two-sided black hole of mass $M$.  The question above, therefore, 
can be answered by evolving $\ketc{\Psi_0}$ in Eq.~(\ref{eq:Psi_0}) 
with respect to time $v$ in the non-null Kruskal-Szekeres 
coordinates:
\begin{equation}
  \left\{ \begin{array}{l} 
    u = \frac{1}{2}(-U+V), \\
    v = \frac{1}{2}(U+V), \end{array} \right.
\qquad
  \left\{ \begin{array}{l} 
    U = -R e^{-\tau}, \\
    V = R e^{\tau}, \end{array} \right.
\label{eq:Kruskal}
\end{equation}
where $R$ and $\tau$ are given for $r > 2M l_{\rm P}^2$ by
\begin{equation}
  R = Ml_{\rm P}^2 \sqrt{\frac{r}{2Ml_{\rm P}^2}-1}\, 
    e^{\frac{r}{4Ml_{\rm P}^2}},
\qquad
  \tau = \frac{t-t_*}{4Ml_{\rm P}^2}.
\label{eq:R-tau}
\end{equation}
Specifically, the original annihilation and creation operators, 
$b_\gamma$, $b_\gamma^\dagger$, $\tilde{b}_\gamma$, and 
$\tilde{b}_\gamma^\dagger$, can be related with the new 
annihilation and creation operators by
\begin{align}
  a_\xi &= \sum_\gamma 
    \bigl( \alpha_{\xi\gamma} b_\gamma 
    + \beta_{\xi\gamma} b_\gamma^\dagger 
    + \zeta_{\xi\gamma} \tilde{b}_\gamma 
    + \eta_{\xi\gamma} \tilde{b}_\gamma^\dagger \bigr),
\label{eq:a_xi}\\
  a_\xi^\dagger &= \sum_\gamma 
    \bigl( \beta_{\xi\gamma}^* b_\gamma 
    + \alpha_{\xi\gamma}^* b_\gamma^\dagger 
    + \eta_{\xi\gamma}^* \tilde{b}_\gamma 
    + \zeta_{\xi\gamma}^* \tilde{b}_\gamma^\dagger \bigr),
\label{eq:a_xi-dag}
\end{align}
where $\xi$ is the label for modes in which the frequency $\omega$ 
with respect to $t$ is replaced by the frequency $\Omega$ with 
respect to $v$, i.e.\ $\xi = \{ \Omega, l, m, \cdots \}$, and 
$\alpha_{\xi\gamma}$, $\beta_{\xi\gamma}$, $\zeta_{\xi\gamma}$, 
and $\eta_{\xi\gamma}$ are the Bogoliubov coefficients calculable 
using the standard quantum field theory method.  The time evolution 
operator in $v$ is then given by
\begin{equation}
  H_v = \sum_\xi \Omega a_\xi^\dagger a_\xi 
    + H_{\rm int}\bigl( a_\xi, a_\xi^\dagger \bigr).
\label{eq:H_v}
\end{equation}
The resulting physics is that of a smooth horizon with interior 
spacetime.

We stress that the ``thermal radiation'' in Eq.~(\ref{eq:rho_hard}), 
obtained by tracing out soft (and far) modes, is very different 
from ``real radiation'' emitted from normal matter, e.g.\ a piece 
of coal, which does not admit a similar construction.%
\footnote{I thank Raphael Bousso for asking a question that has 
 led me to make the comment here.}
In the case of the black hole thermal atmosphere, the form of the 
density matrix---or temperature---is {\it universal} throughout 
the species, reflecting the fact that its thermal nature arises 
from entanglement between the hard and soft modes for each species. 
On the other hand, in the case of radiation from normal matter, 
the structure of the radiation depends on dynamics.  For example, 
depending on couplings between the constituents of matter and 
radiation, it is possible that some species (e.g.\ photons) are 
radiated but not others (e.g.\ neutrinos).  The structure of 
entanglement also depends on the system:\ different configurations 
of radiation are purified by different microscopic configurations 
of matter.  This implies that hard/far---or semiclassical---modes 
comprising the radiation are purified by other semiclassical modes; 
in the language of holography, the purifier of radiation states 
can be found in code subspace degrees of freedom that allow for 
subsystem recovery.  The universality discussed here is an important 
ingredient for the purifier to be interpreted as comprising 
spacetime, which occurs for the black hole thermal atmosphere 
and Unruh radiation (see Section~\ref{subsec:sc-Rindler}).%
\footnote{This reveals an intriguing relation between ultraviolet 
 and infrared physics:\ in order to have spacetime behind the 
 horizon, dynamics at the stretched horizon---i.e., at the string 
 scale $1/l_{\rm s}$---must be chaotic {\it across all low-energy 
 species}.  Specifically, it redistributes the energy of matter 
 falling into the stretched horizon universally among the species. 
 This provides nontrivial information about the dynamics at the scale 
 $1/l_{\rm s}$.  In particular, it must not have a structure preventing 
 the universal redistribution, such as an exact global symmetry.}

We also emphasize that because of an extremely large boost between 
distant and infalling reference frames, the evolution in the 
black hole interior generated by Eq.~(\ref{eq:H_v}) occurs 
``instantaneously'' from the viewpoint of a distant frame, i.e.\ 
within a cutoff time as measured locally at $r \approx r_{\rm s}$ 
(within $\approx M l_{\rm P}^2$ in Schwarzschild time). 
This implies that it is not possible to manipulate soft 
and far (radiation) modes to affect a mirror state---the state 
associated with a given hard mode state $\ket{\{ n_\alpha \}}$ 
in Eq.~(\ref{eq:BH-occup})---within the time scale relevant for 
the effective description.  This provides a further justification 
for the coarse-graining in Eq.~(\ref{eq:coarse}) and is a key to 
understand the apparent uniqueness of the infalling vacuum, despite 
the existence of exponentially many black hole microstates.

There is an important restriction on the applicability of the 
effective description discussed above.  Since $\ket{\{ n_\alpha \}}$ 
represents states of the hard modes, i.e.\ the semiclassical 
modes in the zone $r_{\rm s} \leq r \leq r_{\rm z}$, at $t = t_*$, 
$\ketc{\{ n_\alpha \}}$ represents states of their mirror modes, 
i.e.\ the semiclassical modes in the corresponding mirror region 
in the second exterior of the two-sided black hole, at the 
time when the vacuum state takes the thermofield double form, 
Eq.~(\ref{eq:BH-coarse}) (which we also denote by $t_*$).  The 
effective description obtained by the coarse-graining, therefore, 
is well-defined only in the domain of dependence, $D_{\rm z}$, of 
the union of the zone and its mirror regions at $t = t_*$ in the 
two-sided description.  This implies that a given effective theory 
defined by $\ketc{\Psi_0}$ and $H_v$ may describe only a part of 
the history of a falling object.

\begin{figure}[t]
\begin{center}
  \includegraphics[height=6.5cm]{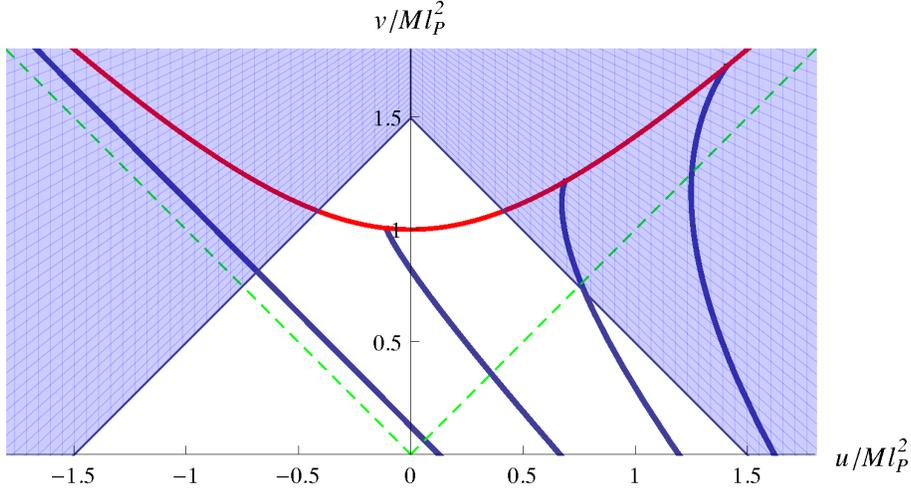}
\end{center}
\caption{Trajectories of an object released from $r = 4M l_{\rm P}^2$ 
 at $(t - t_*)/Ml_{\rm P}^2 = -8, -10, -13, -20$, depicted in the 
 non-null Kruskal-Szekeres coordinates $(u,v)$ (solid dark-blue 
 lines, from right to left).  The unfilled triangle-shaped region 
 represents the positive $v$ part of the spacetime described by 
 the effective theory of the interior built on the full microstate 
 at $t = t_*$.  The solid red line and dashed green 
 lines represent the singularity, $v = \sqrt{u^2 + M^2 l_{\rm P}^2}$, 
 and the horizon, $v = |u|$, respectively.}
\label{fig:traj}
\end{figure}

To illustrate this point, we have depicted in Fig.~\ref{fig:traj} 
the trajectories of an object released from $r = 4M l_{\rm P}^2$ at 
$(t - t_*)/Ml_{\rm P}^2 = -8, -10, -13, -20$ in the $u$-$v$ plane 
(from right to left). 
The unfilled triangle is the spacetime region in $v \geq 0$ that 
can be described by this effective theory, which is determined by 
the location of the edge of the zone at $t = t_*$:\ $(u,v) \simeq 
(1.5 Ml_{\rm P}^2, 0)$.  Note that light rays travel at $45^\circ$ 
in the $u$-$v$ plane, since
\begin{equation}
  ds^2 = \frac{32Ml_{\rm P}^2}{r(u,v)} 
    e^{-\frac{r(u,v)}{2Ml_{\rm P}^2}} 
    (-dv^2 + du^2) + r(u,v)^2 d\Omega^2.
\label{eq:met-uv}
\end{equation}
We find that the trajectory of the object released at 
$t = t_* - 13 Ml_{\rm P}^2$ can be described until it hits the 
singularity,%
\footnote{By an object hitting the singularity, we mean the 
 object entering the region near $r = 0$ in which the semiclassical 
 description of gravity breaks down, specifically the region in 
 which a curvature invariant exceeds the string scale, $r \lesssim 
 (l_{\rm s}^2 l_{\rm P}^2 M)^{1/3}$.}
while other trajectories cannot.  To fully describe other 
trajectories, we need to erect different effective theories 
building on states at different times.  For example, to describe 
the trajectory of the object released from $r = 4M l_{\rm P}^2$ 
at $t = t_* - 8 Ml_{\rm P}^2$, we can build an effective theory 
on the state at $t = t_* + 5 Ml_{\rm P}^2$
\begin{equation}
  \ket{\Psi} = e^{-iH(5 Ml_{\rm P}^2)} \ket{\Psi(t=t_*)},
\label{eq:t-5M}
\end{equation}
where $H$ is the time evolution operator in Schwarzschild time $t$.

The existence of a consistent semiclassical description based on 
the $v$ evolution implies that there is a subsector in the original 
microscopic theory in which the physics perceived by an object 
after it crosses the horizon can be described unitarily until it 
hits the singularity or leaves the spacetime region given by the 
effective theory.  Note that we can always find an effective theory 
describing the full history of an object.  As can be seen from 
Fig.~\ref{fig:traj}, if we erect an effective theory sufficiently 
early, the object leaves $D_{\rm z}$ in the positive $u$ direction 
before it hits the singularity.  On the other hand, if the time 
to erect the effective theory is late, then the object leaves the 
region in the negative $u$ direction before reaching the singularity. 
Since the singularity is always located in the spacetime region 
described by an effective theory, continuity tells us that we can 
choose a time to erect the effective theory such that the full 
trajectory of the object is described until it hits the singularity.

So far, we have only considered the infrared cutoff of the effective 
description provided by the end of the zone, $r \lesssim r_{\rm z}$ 
at $t = t_*$.  However, it is also important to consider the 
ultraviolet cutoff given by the stretched horizon, $r \geq r_{\rm s}$ 
at $t = t_*$.%
\footnote{Note that in the effective two-sided description, this 
 excludes the region whose proper distance from the bifurcation 
 surface (at $t = t_*$) is smaller than the string length, i.e.\ 
 the union of the region between the mathematical and stretched 
 horizons, $2Ml_{\rm P}^2 \leq r < r_{\rm s}$, and its corresponding 
 mirror region in the second exterior.}
If there were no such cutoff, as in the case of the classical 
description, then all the matter that fell into the black hole earlier 
than $t \simeq t_* + 4Ml_{\rm P}^2 \ln(Ml_{\rm P}^2/l_{\rm s})$ (the 
intersection of the stretched horizon and the future boundary of 
$D_{\rm z}$) would appear in the effective description, with the 
trajectories of all the objects that fell earlier than $\approx t_*$ 
concentrated near $v = -u$.  This would cause large backreaction on 
the spacetime, destroying the validity of the effective description. 
The existence of the ultraviolet cutoff, however, saves the picture.

To see the implications of the ultraviolet cutoff, let us consider 
an effective theory erected at $t = t_*$ and a process in which a 
falling object sends a null signal in the positive $u$ direction 
just after passing the horizon.  Suppose that the object is at 
$r = r_{\rm s}$ at $t = t_*$, the location closest to the horizon 
with the ultraviolet cutoff.  The signal then leaves $D_{\rm z}$ 
of this effective theory toward positive $u$, but the same signal 
may also appear in another effective theory erected later at 
$t = t_* + \varDelta t$.  In order for this to happen, the signal 
must be sent when
\begin{equation}
  U < R_{\rm z} e^{-\frac{t_*+\varDelta t}{4Ml_{\rm P}^2}},
\label{eq:U-cond-1}
\end{equation}
where $U$ is given by Eq.~(\ref{eq:Kruskal}) and
\begin{equation}
  R_{\rm z} = Ml_{\rm P}^2 \sqrt{\frac{r_{\rm z}}{2Ml_{\rm P}^2}-1}\, 
    e^{\frac{r_{\rm z}}{4Ml_{\rm P}^2}} 
  \simeq 1.5 Ml_{\rm P}^2.
\label{eq:R_z}
\end{equation}
Now, in order for the object to send a nontrivial signal, a proper 
time of order $l_{\rm P}$ must elapse after it passes the horizon. 
This is achieved at the smallest $U$ if the object is dropped at 
$r = r_{\rm s}$ (at $t = t_*$) with zero initial velocity.  However, 
this still requires
\begin{equation}
  U > O(l_{\rm P})\, e^{-\frac{t_*}{4Ml_{\rm P}^2}},
\label{eq:U-cond-2}
\end{equation}
for the signal to be sent, where we have taken $l_{\rm P} \approx 
l_{\rm s}$ anticipating the level of precision in the final result. 
By requiring consistency between Eq.~(\ref{eq:U-cond-1}) and 
Eq.~(\ref{eq:U-cond-2}), we obtain
\begin{equation}
  \varDelta t < 4Ml_{\rm P}^2 \left[ \ln(Ml_{\rm P}) + O(1) \right].
\label{eq:del-t_t0}
\end{equation}
This implies that the effective theory erected at $t$ cannot receive 
any signal sent before $\approx t - 4Ml_{\rm P}^2 \ln(Ml_{\rm P})$ 
inside the horizon.

This strongly suggests that the effective theory erected at $t = t_*$ 
should not include an object that has reached the stretched horizon 
at
\begin{equation}
  t < t_* - 4Ml_{\rm P}^2 \left[ \ln(Ml_{\rm P}) + O(1) \right].
\label{eq:UV-cutoff}
\end{equation}
Note that in the distant description, an object that has reached the 
stretched horizon before $t_*$ appears as excitations of the stretched 
horizon modes at $t = t_*$.  The existence of these modes, together 
with the fact that an excitation of a hard mode can always be 
discriminated from the thermal atmosphere, allows us to avoid 
the frozen vacuum argument in Ref.~\cite{Bousso:2013ifa}.  It 
is then natural to associate the cutoff of Eq.~(\ref{eq:UV-cutoff}) 
with the fact that excitations of the stretched horizon modes are 
eventually dissipated into soft modes.  Specifically, the excitations 
of the stretched horizon modes caused by an object reached at the 
stretched horizon before $t_* - 4Ml_{\rm P}^2 \ln(Ml_{\rm P})$ 
have already resolved into the soft modes by the time the effective 
theory is erected at $t_*$.  This implies that the timescale for 
excitations of the stretched horizon to relax into soft modes is
\begin{equation}
  t_{\rm rel} = 4Ml_{\rm P}^2 \ln(Ml_{\rm P}),
\label{eq:t_rel}
\end{equation}
up to terms that are not enhanced by $\ln(Ml_{\rm P})$.  The agreement 
between this timescale and that in Eq.~(\ref{eq:t_eq}) is suggestive.

Incidentally, the condition of Eq.~(\ref{eq:UV-cutoff}) is identical 
to the condition that an object must enter $D_{\rm z}$ of an 
effective theory before hitting the stretched horizon in order 
for it to be described by the effective theory.  This coincidence 
is comfortable and strengthens our confidence in the validity of 
the cutoff given in Eq.~(\ref{eq:UV-cutoff}).  With this cutoff, 
a small object released from $r = r_0$ at $t = t_0$ with $r_0 - 
2Ml_{\rm P}^2 \approx O(Ml_{\rm P}^2)$ is included in the effective 
theory erected at $t_*$ only if
\begin{equation}
  t_0 > t_* -8Ml_{\rm P}^2 \left[ \ln(Ml_{\rm P}) + O(1) \right].
\label{eq:t0-cond}
\end{equation}
This implies that only the object that is dropped at sufficiently 
late time appears in the effective description.  This makes it clear 
that the issue of large backreaction is avoided.

\begin{figure}[t]
\begin{center}
  \includegraphics[height=6.5cm]{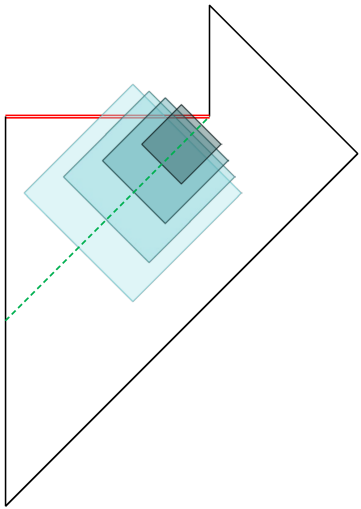}
\end{center}
\caption{A series of effective theories erected at different times 
 (depicted by diamonds) covering the interior spacetime.  The double 
 red line and the dashed green line represent the singularity and the 
 horizon, respectively.  The figure is only a sketch; in particular, 
 the second exterior and the white hole region in each effective 
 theory do not belong to the original spacetime.}
\label{fig:global}
\end{figure}

The discussion above implies that to describe the interior 
of a black hole ``throughout its history,'' one needs to use 
multiple effective theories erected at different times (which 
are generally not mutually independent).  The picture of the 
classical interior spacetime in general relativity emerges only 
after ``patching'' descriptions given by these effective theories; 
see Fig.~\ref{fig:global} for a schematic depiction.  This is how 
the picture of complementarity~\cite{Susskind:1993if} is realized 
in the present framework.  A similar idea was discussed in the 
context of multiverse cosmology in Ref.~\cite{Nomura:2011dt}.

\subsubsection{Resolution of the cloning paradox}
\label{subsubsec:cloning}

A potential issue in a theory in which a black hole evolves unitarily 
is that of cloning of quantum information~\cite{JP,Susskind:1993mu}. 
Suppose that an observer falling into a black hole sends some 
quantum information along the horizon right after he/she passes 
the horizon.  Suppose also that another observer hovering outside 
the horizon decodes this information from Hawking radiation (which 
is possible if the black hole evolution is unitary) and then jumps 
into the black hole afterward.  Now, if the second observer can 
also receive the signal directly from the first observer after 
passing the horizon, then it would mean that the second observer 
has obtained two copies of the same quantum information, which is 
prohibited by linearity of quantum mechanics~\cite{Wootters:1982zz}.

It is often said that this problem is avoided because no one can 
operationally obtain two copies of information, either because 
of the time it takes for a black hole to process and send back 
information~\cite{Hayden:2007cs,Susskind:1993mu} or an exponentially 
long time needed for an observer to decode information from 
Hawking radiation~\cite{Harlow:2013tf}.  The framework discussed 
here, however, provides an arguably simpler solution:\ there is 
no duplicate information in any single description, regardless 
of whether it can be operationally possessed by an observer or 
not.  In this section, we describe this picture.  A similar idea 
has also been discussed in Ref.~\cite{Maldacena:2017axo}, although 
the detailed implementations are different.

First, it is clear that no cloning occurs in the distant description 
because there is no interior.  The issue, therefore, is if any 
effective description of the interior may contain duplicate 
information.  We now argue that the answer is no.  In order for 
the paradox to occur, there must be an infalling object as well 
as radiation from the black hole that contains the same information. 
In the effective description, however, there is no Hawking radiation 
emitted from the edge of the zone, since the region outside the 
zone is not contained in the region described by the effective 
theory.  The thermal atmosphere of the black hole is also absent 
in this description, since the soft modes are already coarse-grained 
to give the semiclassical modes in the mirror zone region.  There 
is simply no way to have information in radiation---or, in fact, 
radiation itself---in the effective theory describing the interior!

This only leaves the following possibility for information 
duplication.  A detector located in the zone retrieves information 
about a fallen object and converts it into information in 
semiclassical degrees of freedom before an effective theory 
describing the falling object in the interior is erected, making 
both the object and converted information appear in the effective 
theory.  This is, however, impossible.  As we have seen in 
Section~\ref{subsubsec:inside}, an effective theory erected 
at $t = t_*$ does not describe an object that has reached the 
stretched horizon before entering the spacetime region $D_{\rm z}$. 
This implies that an object must reach the stretched horizon when
\begin{equation}
  t \geq t_* - 4Ml_{\rm P}^2 \left[ \ln(Ml_{\rm P}) + O(1) \right] 
    \equiv t_1,
\label{eq:t_1}
\end{equation}
in order for it to be included in the description.  Here, we have 
kept explicit only the $\ln(Ml_{\rm P})$ enhanced piece, regarding 
$\ln(l_{\rm s}/l_{\rm P}) \approx O(1)$.  On the other hand, the 
argument below Eq.~(\ref{eq:UV-cutoff}) implies that information 
about an object fallen after $t_1$ stays as excitations of the 
stretched horizon modes until $t_*$, making it impossible for 
a detector to extract it from the soft modes.

The discussion above is sufficient to argue that no information 
duplication occurs in any single description.  One might, however, 
be more satisfied if the information about a falling object 
described by an effective theory can never reappear in the 
spacetime region, $D_{\rm z}$, of the effective theory even when 
the system is described in a distant reference frame.  Note that 
in a distant description, a physical detector may mine information 
from the black hole thermal atmosphere if it is held in the zone 
for a sufficiently long time; see Section~\ref{subsec:mining}.%
\footnote{A detector click also occurs in the effective description, 
 which is associated with an emission of a particle rather than an 
 absorption of a particle in the thermal bath~\cite{Unruh:1983ms}. 
 This, however, does not allow for extracting information from the 
 vacuum state, since the description is already coarse-grained and 
 hence is intrinsically semiclassical.}
This stronger condition eliminates the possibility that a distant 
description finds reappearance of the information within $D_{\rm z}$ 
while the effective description does not.  Such a discrepancy 
between the descriptions is not a contradiction because the 
effective theory involves coarse-graining and hence may lead 
to information loss, precisely as in the usual semiclassical 
theory of gravity.  Nevertheless, it is aesthetically appealing 
if this discrepancy never occurs.  Below we derive a consequence 
of requiring this aesthetic criterion and see if it is reasonable 
to expect that it is satisfied.

The strongest restriction from the requirement arises if an 
effective theory is erected such that the object reaches the 
stretched horizon at $t = t_1$, given in Eq.~(\ref{eq:t_1}). 
The requirement that the information about this objects does 
not reappear in the zone within $D_{\rm z}$ implies that it 
must not reappear from the black hole before
\begin{equation}
  t_2 = t_* + 4Ml_{\rm P}^2 \left[ \ln(Ml_{\rm P}) + O(1) \right],
\label{eq:t_2}
\end{equation}
the time at which the entirety of the outside region, $r > r_{\rm s}$, 
leaves $D_{\rm z}$.  Since what leaves $D_{\rm z}$ last is the 
region near the stretched horizon, $r \approx r_{\rm s}$, this 
condition is equivalent to saying that the information about 
the fallen object should not reappear before $t = t_2$ so that 
even a detector located near the stretched horizon may not probe 
it.  Combining with Eq.~(\ref{eq:t_1}), we can then conclude that 
the black hole must retain information longer than $t_2-t_1 \approx 
8Ml_{\rm P}^2 \ln(Ml_{\rm P}) + O(Ml_{\rm P}^2)$.

Summarizing, the requirement of no information recovery in 
$D_{\rm z}$ implies that the information retention time, 
$t_{\rm I}$, of a black hole of mass $M$ must satisfy
\begin{equation}
  t_{\rm I} \geq 8Ml_{\rm P}^2 \ln(Ml_{\rm P}),
\label{eq:t_I}
\end{equation}
up to terms that are not enhanced by $\ln(Ml_{\rm P})$.  An 
intriguing point is that the coefficient of the log-enhanced 
term, i.e.\ $8$, is determined.  In terms of the temperature, 
$T_{\rm H}$, and entropy, $S_{\rm BH}$, of the black hole, this 
can be written as
\begin{equation}
  t_{\rm I} \geq \frac{1}{2\pi T_{\rm H}} \ln S_{\rm BH} 
  = \frac{1}{\lambda_{\rm L, max}} \ln S_{\rm BH},
\label{eq:t_I-2}
\end{equation}
where $\lambda_{\rm L, max}$ is the upper bound on a Lyapunov 
exponent found in Ref.~\cite{Maldacena:2015waa}, which a black hole 
is expected to saturate.  This expression makes it natural to expect 
that the condition in Eq.~(\ref{eq:t_I}) is indeed satisfied, with 
the inequality saturated up to non-log-enhanced terms.

\section{Rindler Limit}
\label{sec:Rindler}

In this section, we consider the Rindler limit, aiming to clarify 
some confusion in the literature regarding the relation between 
Hawking emission and the Unruh effect.  We first discuss black 
hole mining, which directly corresponds to the Unruh effect in 
the Rindler limit.  We then discuss how physics in Minkowski space 
arises in this limit, especially focusing on the flow of information.

\subsection{Black hole mining}
\label{subsec:mining}

The energy and entropy of a black hole can be extracted directly 
by placing a probe material into the zone~\cite{Unruh:1982ic,%
Brown:2012un}.  This process, called black hole mining, can 
accelerate the extraction of black hole energy and entropy 
compared with Hawking emission.

An important difference between mining and Hawking emission 
processes is the energy cost of angular momentum relative to the 
local temperature of the thermal atmosphere, $T_{\rm loc}(r)$ 
in Eq.~(\ref{eq:s-density}).  A similar estimate as in 
Section~\ref{subsec:falling} tells us that a particle with 
angular momentum $L^2 = l(l+1)$ costs the energy, as measured 
in the asymptotic region, of
\begin{equation}
  \varDelta \omega \approx O\biggl(\frac{l}{r}\biggr),
\label{eq:ang-cost}
\end{equation}
so that
\begin{equation}
  \frac{\varDelta \omega}{T_{\rm loc}(r)} 
  \approx O\Biggl(\frac{l}{T_{\rm H} r} 
    \sqrt{1-\frac{2Ml_{\rm P}^2}{r}}\Biggr) 
  \approx O\Biggl(l \sqrt{1-\frac{2Ml_{\rm P}^2}{r}}\Biggr).
\label{eq:ang-TH}
\end{equation}
Therefore, modes up to $l \approx O(\sqrt{Ml_{\rm P}^2/\delta})$ 
respond to the thermal atmosphere effectively, where we have taken 
$r = 2Ml_{\rm P}^2 + \delta$ ($\delta \ll O(Ml_{\rm P}^2)$).  This 
implies that we can, a priori, utilize many modes
\begin{equation}
  \sum_{l=0}^{l_{\rm max}}(2l+1) \approx l_{\rm max}^2 
  \approx O\biggl( \frac{Ml_{\rm P}^2}{\delta} \biggr),
\label{eq:num-modes}
\end{equation}
to extract black hole energy and entropy, compared with $O(1)$ 
(mostly the $s$-wave mode) in the case of Hawking emission.  At the 
stretched horizon, $\delta \approx l_{\rm s}^2/Ml_{\rm P}^2$, so that 
this number is enormous, $\approx O((Ml_{\rm P}^2/l_{\rm s})^2)$.

In realistic situations, energy conditions applied to the 
probe material prevent us from utilizing all the modes in 
Eq.~(\ref{eq:num-modes}) for $\delta \lesssim O(l_{\rm P})$; 
specifically, the null energy condition enforces the number of 
modes that can be used for mining to be of $O(M l_{\rm P})$ or 
smaller, so that the black hole lifetime cannot be shorter than 
$O(M^2 l_{\rm P}^3)$~\cite{Brown:2012un}.  An important point here, 
however, is that the rate of extracting energy and entropy for 
each mode is the same as that in Hawking emission---the acceleration 
of extraction occurs not because of a higher rate per mode but 
because of an increased number of modes available to the probe 
immersed into the zone.  This justifies the analysis in 
Section~\ref{subsubsec:cloning}, which examines the reappearance 
of information near the stretched horizon to constrain the 
black hole information retention time.

As in the case of Hawking emission, backreaction of mining causes 
ingoing negative energy-entropy excitations.  A difference is that 
in the case of mining, these excitations are generally localized 
in the angular directions.  It is expected that the excitations 
are scrambled in the soft modes at a timescale not much larger than 
$O(M l_{\rm P}^2 \ln(M l_{\rm P}))$.

\subsection{Semiclassical description in Rindler space}
\label{subsec:sc-Rindler}

Here we discuss issues associated with the Rindler limit.  We mostly 
focus on how the semiclassical description of Rindler space is 
related with that of a black hole.  In Section~\ref{subsubsec:info}, 
we extend the comparison beyond the purely semiclassical regime.

\subsubsection{Unruh effect}
\label{subsubsec:Unruh}

Rindler space is obtained as the limiting case of Schwarzschild 
spacetime
\begin{equation}
  M \rightarrow \infty
\quad\mbox{and}\quad
  l_{\rm P}, l_{\rm s}: \mbox{ fixed},
\label{eq:Rind-limit}
\end{equation}
by focusing on the near horizon region, $r \rightarrow 
2Ml_{\rm P}^2$, such that the combinations
\begin{equation}
  \rho \equiv 2\sqrt{2Ml_{\rm P}^2(r-2Ml_{\rm P}^2)}
\qquad\mbox{and}\qquad
  \tau \equiv \frac{t}{4Ml_{\rm P}^2},
\label{eq:rho-omega}
\end{equation}
are kept finite.  The metric in this limit is given by the standard 
Rindler form
\begin{equation}
  ds^2 = -\rho^2 d\tau^2 + d\rho^2 + r(\rho)^2 d\Omega^2.
\label{eq:Rindler}
\end{equation}

There is no direct analogue of Hawking emission in the Rindler 
limit, since the edge of the zone in the original Schwarzschild 
spacetime is now at spatial infinity.  (We implicitly imagine 
an infrared cutoff $\rho_{\rm IR} \rightarrow \infty$ so that 
$\rho_{\rm IR}/Ml_{\rm P}^2 < \infty$.)  There is, however, an 
analogue of the mining process with a physical probe sensing 
a thermal bath with temperature
\begin{equation}
  T_{\rm loc}(\rho) = \lim_{M \rightarrow \infty} 
    \frac{T_{\rm H}}{\sqrt{1-\frac{2Ml_{\rm P}^2}{r}}} 
  = \frac{1}{2\pi \rho}.
\label{eq:Rindler-T}
\end{equation}
This is the well-known Unruh effect~\cite{Unruh:1976db,%
Fulling:1972md,Davies:1974th}.  In the black hole case, mining 
allows us to extract information about a black hole vacuum. 
(Recall that a black hole background appears as a black hole 
vacuum at the semiclassical level, although it actually represents 
a collection of black hole microstates.)  On the other hand, 
we do not expect to extract information about the Minkowski 
vacuum from Unruh radiation.  Is there a fundamental 
difference between black hole mining and the Unruh effect, 
beyond the fact that the latter requires the limit of 
Eqs.~(\ref{eq:Rind-limit},~\ref{eq:rho-omega}) to be taken?

There isn't.  In order for information about a black hole 
vacuum to be extracted, more than a half of its entropy must 
be mined~\cite{Page:1993wv}.  The entropy of the Minkowski vacuum, 
however, is infinite
\begin{equation}
  S_{\rm Minkowski} = \infty,
\label{eq:S_Minkowski}
\end{equation}
as can be seen from the fact that it is obtained by taking the 
$M l_{\rm P} \rightarrow \infty$ limit of Schwarzschild spacetime 
(or the $H l_{\rm P} \rightarrow 0$ limit of de~Sitter space, 
where $H$ is the Hubble parameter).  Therefore, no finite size 
detector can collect more than a half of the entropy in any finite 
time.  The process of quickly recovering newly added information 
considered in Ref.~\cite{Hayden:2007cs} is not available either, 
because Minkowski space cannot be maximally entangled with any 
finite system (and because the ``scrambling time'' of Rindler 
space is infinite; see Section~\ref{subsubsec:info}).  Information 
about the Minkowski vacuum, thus, cannot be mined using the Unruh 
effect by any physical detector---detecting Unruh radiation simply 
corresponds to an infinitesimally early portion of the Page curve.%
\footnote{The fact that Rindler space corresponds to an infinitely 
 young black hole implies that the smoothness of Minkowski space 
 cannot by itself be used to argue against the firewall phenomenon 
 discussed in Refs.~\cite{Almheiri:2012rt,Almheiri:2013hfa,%
 Marolf:2013dba}.}
This is related to the statement that the Bondi-Metzner-Sachs 
(BMS)~\cite{Bondi:1962px,Sachs:1962wk} soft charges, representing 
microstates of the Minkowski vacuum, cannot be measured 
by experiments in finite time using a finite-size detector, 
whose results are determined by the conventional 
$S$-matrix~\cite{Mirbabayi:2016axw,Bousso:2017dny}.

As in the case of mining a young black hole, detecting Unruh 
radiation generates entanglement between the detector and the modes 
represented by the thermal bath, i.e.\ the soft modes, through 
the creation of localized negative energy-entropy excitations. 
This can be understood from the fact that in a Minkowski frame, 
the Unruh effect corresponds to emission of particles from the 
detector~\cite{Unruh:1983ms}, which generates entanglement between 
them.  The generated entanglement, however, is finite, so it is 
infinitesimally small compared with the infinite amount of entropy 
the soft modes have.

\subsubsection{Mirror operators}
\label{subsubsec:mirror}

The description of Rindler space is analogous to that of a 
black hole in the zone.  In particular, semiclassical theory 
in Rindler space describes microscopic dynamics of only the 
hard modes, whose locally measured energies are sufficiently 
larger than $T_{\rm loc}(\rho)$.  Following the notation in 
Section~\ref{subsubsec:two-sided}, we denote the states of 
these modes by $\ket{\{ n_\alpha \}}$.  The other degrees of 
freedom are regarded as the soft modes, which can be described 
only statistically.

There is, however, a notable difference arising from taking the 
limit of Eq.~(\ref{eq:Rind-limit},~\ref{eq:rho-omega}):\ the 
state-dependent nature of constructing the ``interior space'' 
(the other side of the Rindler horizon) becomes irrelevant. 
We first note that since there is no early Hawking mode, the 
mirror space is constructed purely out of the soft modes.  This, 
however, does not by itself eliminate the need of state dependence. 
A natural microscopic definition of state-independent mirror 
operators would be
\begin{align}
  \tilde{b}_\gamma 
  &= {\bf 1} \otimes 
    \sqrt{n_\gamma} \sum_{i = 1}^{{\cal N}(M-E_n)} 
    \ket{\psi_i(M-E_{n_-})} \bra{\psi_i(M-E_n)},
\label{eq:m-ann}\\
  \tilde{b}_\gamma^\dagger 
  &= {\bf 1} \otimes 
    \sqrt{n_\gamma + 1} \sum_{i = 1}^{{\cal N}(M-E_{n_+})} 
    \ket{\psi_i(M-E_{n_+})} \bra{\psi_i(M-E_n)},
\label{eq:m-cre}
\end{align}
where $\ket{\psi_i(M-E_n)}$ are the soft mode states associated 
with $\ket{\{ n_\alpha \}}$, and
\begin{equation}
  n_- = \{ n_\alpha - \delta_{\alpha\gamma} \},
\qquad
  n_+ = \{ n_\alpha + \delta_{\alpha\gamma} \}.
\label{eq:n-_n+}
\end{equation}
Here, the unit matrices in Eqs.~(\ref{eq:m-ann},~\ref{eq:m-cre}) 
represent the fact that these operators act trivially on the hard 
modes, and we have kept the ``black hole mass'' $M$, although it 
is taken to infinity in the Rindler limit.

A potential problem lies in the creation operators 
$\tilde{b}_\gamma^\dagger$.  Suppose $M$ were finite. 
The number of independent soft mode microstates associated 
with $\ket{n_+}$ would then be smaller than that of $\ket{n}$, 
which prevents us from defining a microscopic operator that maps 
any state in the space spanned by $\ket{\psi_i(M-E_n)}$ into a 
state in the space spanned by $\ket{\psi_i(M-E_{n_+})}$ preserving 
the appropriate normalization.  The fractional difference between 
the number of independent soft mode states associated with 
$\ket{n_+}$ and $\ket{n}$, however, vanishes in the limit 
$M \rightarrow \infty$ for fixed energies.  The state-independent 
operators in Eqs.~(\ref{eq:m-ann},~\ref{eq:m-cre}) thus serve 
as valid mirror operators in this limit.

\subsubsection{Information retrieval}
\label{subsubsec:info}

We have seen that the Unruh effect in Rindler space does not allow 
us to extract information about the Minkowski vacuum.  Does this 
mean that information on the other side of the Rindler horizon, 
representing a half of Minkowski space, can never be retrieved in 
a Rindler description?

As seen in Section~\ref{subsubsec:Unruh}, a detection of Unruh 
radiation creates entanglement between the detector and localized 
negative energy-entropy excitations of the Rindler soft modes. 
From a Minkowski point of view, this is entanglement between 
the detector and particles emitted from it.  This entanglement 
can be retrieved in a Rindler description if we decelerate the 
detector adiabatically.  In other words, if we slowly decrease 
the acceleration characterizing the Rindler description, then the 
entanglement---information about the other side of the horizon---can 
be retrieved in Rindler space in a way that it can be described 
by semiclassical theory.  The physical picture is that as the 
Rindler horizon recedes due to the deceleration, particles emitted 
earlier from the detector (in a Minkowski point of view) reappear 
from the horizon, which increases purity of the system that can 
be described by semiclassical theory in Rindler space.

This can be viewed as an analogue of information retrieval from 
a black hole.  A difference is that since the entropy of the 
Minkowski vacuum is infinite, Eq.~(\ref{eq:S_Minkowski}), the 
``scrambling time'' of Rindler space is infinite
\begin{equation}
  \tau_{\rm scr} \approx O\bigl(\rho\, \ln S_{\rm Minkowski}\bigr) 
  \rightarrow \infty.
\label{eq:tau_scr}
\end{equation}
In other words, a state having the negative energy-entropy 
excitations does not relax into a vacuum state in any finite 
time.  This is the reason why the system reappearing from the 
horizon (particles emitted earlier from the detector) is not 
thermalized when it is retrieved.

\section{Entangled Black Holes}
\label{sec:pair}

In this section, we discuss entangled black holes.  As in the earlier 
sections, we consider Schwarzschild (or small AdS) black holes. 
We find that their physics is different from that of commonly 
considered entangled large AdS black holes in a thermal state.

\subsection{A pair of black holes}
\label{subsec:BH-pair}

We first consider a pair of entangled black holes.  We are not 
concerned about how it is actually formed.  It may, for example, 
be formed by preparing many Einstein-Podolsky-Rosen (EPR) pairs 
and collapsing them into a black hole in each side of the pairs. 
Throughout, we assume that each black hole has mass $M$ within 
precision $\varDelta M$.

\subsubsection{Entanglement structure}
\label{subsubsec:pair-ent}

The most general entanglement structure involving the two black 
holes is
\begin{equation}
  \ket{\psi(M,M)} = \sum_{E,F} \sum_{i_E = 1}^{{\cal N}(M-E)} 
    \sum_{j_F = 1}^{{\cal N}(M-F)} \sum_a c_{E i_E F j_F a} 
    \ket{E}_1 \ket{\psi_{i_E}(M-E)}_1 
    \ket{F}_2 \ket{\psi_{j_F}(M-F)}_2 \ket{r_a},
\label{eq:BH-pair-gen}
\end{equation}
where $\sum_{E,F} \sum_{i_E = 1}^{{\cal N}(M-E)} 
\sum_{j_F = 1}^{{\cal N}(M-F)} \sum_a |c_{E i_E F j_F a}|^2 = 1$. 
The subscripts $1$ and $2$ in the states indicate that they are 
states of the first and second black holes, respectively, and 
$\ket{r_a}$ represents states outside the zones of the two black 
holes, including Hawking radiation emitted earlier.

Tracing out the second black hole in Eq.~(\ref{eq:BH-pair-gen}), 
we obtain
\begin{align}
  \rho_{1 {\rm R}}(M) &= {\rm Tr}_2 \ket{\psi(M,M)} \bra{\psi(M,M)} 
\nonumber\\
  &= \sum_{E,E'} \sum_{i_E = 1}^{{\cal N}(M-E)} 
    \sum_{i'_{E'} = 1}^{{\cal N}(M-E')} \sum_{a,a'} 
    d_{E i_E a E' i'_{E'} a'} \ket{E}_1 \ket{\psi_{i_E}(M-E)}_1 
    \ket{r_a}\, {}_1\bra{E'} {}_1\bra{\psi_{i'_{E'}}(M-E')} 
    \bra{r_{a'}},
\label{eq:rho_1}
\end{align}
where
\begin{equation}
  d_{E i_E a E' i'_{E'} a'} = \sum_F \sum_{j_F = 1}^{{\cal N}(M-F)} 
    c_{E i_E F j_F a} c_{E' i'_{E'} F j_F a'}^*.
\label{eq:c_EE'}
\end{equation}
This is different from the state of the combined system of a 
non-entangled black hole and radiation in Eq.~(\ref{eq:BH-rad}), 
i.e.\ $\rho_{1R}(M) \neq \ket{\psi(M)} \bra{\psi(M)}$, unless 
the system associated with the second black hole decouples, 
$c_{E i_E F j_F a} = c^{(1)}_{E i_E a_1} c^{(2)}_{F j_F a_2}$ with 
$\ket{r_a} \rightarrow \ket{r^{(1)}_{a_1}} \ket{r^{(2)}_{a_2}}$. 
Nevertheless, the reduced density matrix for the hard modes obtained 
from Eq.~(\ref{eq:rho_1})
\begin{align}
  \rho_{{\rm H}_1}(M) &= {\rm Tr}_{{\rm S}_1 {\rm R}}\, 
    \rho_{1 {\rm R}}(M) 
\nonumber\\
  &= \sum_E \sum_{i_E = 1}^{{\cal N}(M-E)} \sum_a 
    d_{E i_E a E i_E a} \ket{E}_1\, {}_1\bra{E} 
\nonumber\\
  &\simeq \frac{1}{\sum_E e^{-\frac{E}{T_{\rm H}}}} 
    \sum_E e^{-\frac{E}{T_{\rm H}}} \ket{E}_1\, {}_1\bra{E},
\label{eq:rho_H1}
\end{align}
takes the standard thermal form, as in Eq.~(\ref{eq:rho_hard}). 
Thus, the physics at the semiclassical level is identical between 
non-entangled and entangled black holes.  Obviously, the same 
applies to the second black hole as well.

Let us now see the correlation between the hard modes of the two 
black holes.  Tracing out the radiation and soft mode states, we 
obtain
\begin{align}
  \rho_{{\rm H}_1 {\rm H}_2}(M,M) 
  &= {\rm Tr}_{{\rm S}_1 {\rm S}_2 {\rm R}} 
    \ket{\psi(M,M)} \bra{\psi(M,M)} 
\nonumber\\
  &= \sum_{E,F} \left( \sum_{i_E = 1}^{{\cal N}(M-E)} 
    \sum_{j_F = 1}^{{\cal N}(M-F)} \sum_a 
    |c_{E i_E F j_F a}|^2 \right) 
    \ket{E}_1 \ket{F}_2\, {}_1\bra{E} {}_2\bra{F} 
\nonumber\\
  &\simeq \frac{1}{\sum_{E,F} e^{-\frac{E+F}{T_{\rm H}}}} 
    \left( \sum_E e^{-\frac{E}{T_{\rm H}}} 
      \ket{E}_1\, {}_1\bra{E} \right) \otimes 
    \left( \sum_F e^{-\frac{F}{T_{\rm H}}} 
      \ket{F}_2\, {}_2\bra{F} \right),
\label{eq:rho_H1H2}
\end{align}
where in the last line we have assumed genericity of the 
coefficients $c_{E i_E F j_F a}$.  We find that the hard modes 
of the two black holes are generically not correlated at the 
semiclassical level.  In fact, the lack of quantum entanglement 
between these modes is rather general.  To illustrate it, 
consider the special case that the hard modes of the two black 
holes are correlated at the microscopic level:\ $c_{E i_E F j_F a} 
\propto \delta_{E F}$.  In this case, the state of the entire 
system can be written without loss of generality as
\begin{equation}
  \ket{\psi(M,M)} = \sum_E \sum_{i_E = 1}^{{\cal N}(M-E)} \sum_a 
    c_{E i_E a} \ket{E}_1 \ket{\psi_{i_E}(M-E)}_1 
    \ket{E}_2 \ket{\psi_{i_E}(M-E)}_2 \ket{r_a},
\label{eq:BH-pair}
\end{equation}
where $\sum_E \sum_{i_E = 1}^{{\cal N}(M-E)} \sum_a |c_{E i_E a}|^2 
= 1$.  When the entanglement between the black holes and environment 
is negligible, $c_{E i_E a} \approx c_{E i_E} c_a$, the two black 
holes are strongly entangled with each other at the microscopic 
level; for $c_{E i_E}$'s all having a similar size, they are 
maximally entangled statistically.  Despite this, the correlation 
between the hard modes of the two black holes is given by
\begin{align}
  \rho_{{\rm H}_1 {\rm H}_2}(M,M) 
  &= \sum_E \left( \sum_{i_E = 1}^{{\cal N}(M-E)} \sum_a 
    |c_{E i_E a}|^2 \right) \ket{E}_1 \ket{E}_2\, 
    {}_1\bra{E} {}_2\bra{E} 
\nonumber\\
  &\simeq \frac{1}{\sum_E e^{-\frac{E}{T_{\rm H}}}} 
    \sum_E e^{-\frac{E}{T_{\rm H}}} 
    \ket{E}_1 \ket{E}_2\, {}_1\bra{E} {}_2\bra{E},
\label{eq:rho_H1H2-sp}
\end{align}
so that their correlation is entirely classical.

The origin of the classical nature of the correlation between 
the hard modes is that in our setup, frequencies of the hard modes, 
collectively denoted by $\omega$, are larger than the resolution 
of energy
\begin{equation}
  \omega > \varDelta M.
\label{eq:reason}
\end{equation}
Note that this situation is different from that considered in 
Refs.~\cite{Maldacena:2001kr,Gao:2016bin}, which discuss entangled 
large AdS black holes in a thermal state.

\subsubsection{Interior spacetime without a wormhole}
\label{subsubsec:pair-int}

The fact that the correlation between the hard modes of two entangled 
black holes is at most classical implies that two objects dropped 
into the two black holes cannot meet inside.  Namely, there is no 
wormhole in the sense that objects dropped into different black 
holes can meet in interior spacetime.  Nevertheless, as we will 
see below, each object smoothly enters the horizon of the black 
hole to which it is falling.

As in the case of a non-entangled black hole, the fate of a small 
object dropped, e.g., into the first black hole can be described 
by an effective theory erected around the time when the object 
reaches the stretched horizon.  Assuming that the system is in 
the state of Eq.~(\ref{eq:BH-pair-gen}) at the time the effective 
theory is erected, mirror states in the effective theory are 
given by
\begin{equation}
  \ketc{E}_1 = \frac{1}{\sqrt{z_E}} \sum_{i_E = 1}^{{\cal N}(M-E)} 
    \sum_F \sum_{j_F = 1}^{{\cal N}(M-F)} \sum_a 
    c_{E i_E F j_F a} \ket{\psi_{i_E}(M-E)}_1 
    \ket{F}_2 \ket{\psi_{j_F}(M-F)}_2 \ket{r_a},
\label{eq:mirror-BH1}
\end{equation}
where $z_E = \sum_{i_E = 1}^{{\cal N}(M-E)} \sum_F 
\sum_{j_F = 1}^{{\cal N}(M-F)} \sum_a |c_{E i_E F j_F a}|^2$. 
The (vacuum) state in the effective theory, therefore, takes 
the form
\begin{align}
  \ketc{\psi(M)}_1 &= \sum_E \sqrt{z_E} \ket{E}_1 \ketc{E}_1 
\nonumber\\
  &\simeq \frac{1}{\sqrt{\sum_E e^{-\frac{E}{T_{\rm H}}}}} 
    \sum_E e^{-\frac{E}{2 T_{\rm H}}} \ket{E}_1 \ketc{E}_1,
\label{eq:eff-BH1}
\end{align}
where genericity of the coefficients $c_{E i_E F j_F a}$ is 
assumed in the second line.  The object falling into the first 
black hole can be described by operators analogous to those in 
Eqs.~(\ref{eq:ann},~\ref{eq:cre},~\ref{eq:ann-m},~\ref{eq:cre-m}) 
acting on appropriate factors of Eq.~(\ref{eq:eff-BH1}). The 
resulting physics is that of a smooth horizon for the first 
black hole.

The same is true for an object falling into the second black hole. 
Its fate is described by an effective theory erected around the time 
when the object reaches the stretched horizon of the second black 
hole, now with the identification
\begin{equation}
  \ketc{F}_2 = \frac{1}{\sqrt{z_F}} \sum_E 
    \sum_{i_E = 1}^{{\cal N}(M-E)} \sum_{j_F = 1}^{{\cal N}(M-F)} 
    \sum_a c_{E i_E F j_F a} \ket{E}_1 \ket{\psi_{i_E}(M-E)}_1 
    \ket{\psi_{j_F}(M-F)}_2 \ket{r_a},
\label{eq:mirror-BH2}
\end{equation}
where $z_F = \sum_E \sum_{i_E = 1}^{{\cal N}(M-E)} 
\sum_{j_F = 1}^{{\cal N}(M-F)} \sum_a |c_{E i_E F j_F a}|^2$. 
This leads to
\begin{align}
  \ketc{\psi(M)}_2 &= \sum_F \sqrt{z_F} \ket{F}_2 \ketc{F}_2 
\nonumber\\
  &\simeq \frac{1}{\sqrt{\sum_F e^{-\frac{F}{T_{\rm H}}}}} 
    \sum_F e^{-\frac{F}{2 T_{\rm H}}} \ket{F}_2 \ketc{F}_2.
\label{eq:eff-BH2}
\end{align}
The object falling into the second black hole is described by 
operators analogous to those in Eqs.~(\ref{eq:ann},~\ref{eq:cre},%
~\ref{eq:ann-m},~\ref{eq:cre-m}) acting on this state, which sees 
smooth spacetime when entering the horizon.

\subsection{More than two black holes}
\label{subsec:BH-more}

The analysis described above can be easily generalized to more than 
two black holes with arbitrary masses.  For $n$ black holes with 
masses $M_\alpha$ ($\alpha = 1,\cdots,n$), the general state can 
be written as
\begin{equation}
  \ket{\psi(\{ M_\alpha \})} 
  = \left( \prod_{\alpha = 1}^n\; \sum_{E_\alpha} 
    \sum_{i_{E_\alpha} = 1}^{{\cal N}(M_\alpha-E_\alpha)} 
    \right) \sum_a c_{\{E_\alpha\} \{i_{E_\alpha}\} a} 
    \left( \prod_{\alpha = 1}^n \ket{E_\alpha}_\alpha 
    \ket{\psi_{i_{E_\alpha}}(M_\alpha-E_\alpha)}_\alpha 
    \right) \ket{r_a}.
\label{eq:BH-n-gen}
\end{equation}
We can trace out various components to see the entanglement structure 
of the state.

A general feature of these systems is that the correlations between 
the hard modes of different black holes are at most classical. 
In particular, for generic microscopic entanglement, we find
\begin{align}
  \rho_{\rm H}(\{ M_\alpha \}) 
  &= {\rm Tr}_{{\rm S} {\rm R}} 
    \ket{\psi(\{ M_\alpha \})} \bra{\psi(\{ M_\alpha \})} 
\nonumber\\
  &= \sum_{E_\alpha} \left\{ \left( \prod_{\alpha = 1}^n 
    \sum_{i_{E_\alpha} = 1}^{{\cal N}(M_\alpha-E_\alpha)} \right) 
    \sum_a |c_{\{E_\alpha\} \{i_{E_\alpha}\} a}|^2 \right\} 
    \left( \prod_{\alpha = 1}^n \ket{E}_\alpha\, 
      {}_\alpha\bra{E} \right) 
\nonumber\\
  &\simeq \frac{1}{\sum_{E_\alpha} 
    e^{-\frac{\sum_\alpha E_\alpha}{T_{\rm H}}}} 
    \bigotimes_{\alpha = 1}^n \left( \sum_{E_\alpha} 
      e^{-\frac{E_\alpha}{T_{\rm H}}} 
      \ket{E}_\alpha\, {}_\alpha\bra{E} \right).
\label{eq:rho_Hn}
\end{align}
Small semiclassical objects dropped into different black holes, 
therefore, cannot meet inside.  It is also not possible to make 
the wormhole traversable by connecting hard modes of different 
black holes through small direct interactions.%
\footnote{We can have standard wormhole phenomena between 
 semiclassical objects and ``objects'' that are cleverly composed 
 of hard, soft, and radiation degrees of freedom such that they 
 appear as semiclassical objects dropped from mirror space.}

\section{Conclusions and Discussion}
\label{sec:concl}

In this paper, we have analyzed the quantum mechanics of an 
evaporating black hole.  A key ingredient is that semiclassical 
theory in a distant view (a view based on Schwarzschild time) 
describes microscopic dynamics of only the hard modes:\ the degrees 
of freedom that are hard enough to be discriminated within the 
characteristic timescale of black hole evolution, $t_{\rm H} 
\approx Ml_{\rm P}^2$.  In an equilibrated black hole, these degrees 
of freedom must be entangled with the rest of the black hole degrees 
of freedom, which are too soft to be discriminated due to a large 
redshift caused by the black hole.  This is because the differences 
of the energy between different configurations of the hard modes 
are compensated by those of the soft modes in a black hole microstate, 
whose energy is determined with precision of order $1/t_{\rm H}$. 
In fact, this equilibrium between the hard and soft modes is the 
origin of the thermodynamic nature of a black hole seen at the 
semiclassical level.

Based on intuition coming from comparing usual thermodynamic 
entropies of matter with the Bekenstein-Hawking entropy, we expect 
that the number of the hard mode degrees of freedom is much smaller 
than that of the soft mode degrees of freedom.  This implies that 
the former is much smaller than both the latter and the number of 
early Hawking radiation degrees of freedom
\begin{equation}
  S_{\rm hard} \ll S_{\rm soft}, S_{\rm rad},
\label{eq:S-rel-concl}
\end{equation}
throughout the history of the black hole evolution (except possibly 
in the very earliest time).  Here, $S_{\rm hard}$, $S_{\rm soft}$, 
$S_{\rm rad}$ are the coarse-grained entropies of the hard modes, 
soft modes, and early radiation, respectively.  We have seen that 
this makes the entanglement between these three types of degrees 
of freedom intrinsically tripartite
\begin{equation}
  \ket{\Psi(M)} \sim 
    \sum_E e^{-\frac{E}{2 T_{\rm H}}} 
    \sum_{i_E=1}^{{\cal N}_E} \frac{1}{\sqrt{{\cal N}_E}} 
    \ket{E} \ket{\psi_{i_E}} \ket{r_{i_E}},
\label{eq:GHZ-concl}
\end{equation}
where ${\cal N}_E = {\rm min}\{ e^{S_{{\rm soft},E}}, e^{S_{\rm rad}} 
\}$ with $S_{{\rm soft},E}$ being the coarse-grained entropy of the 
soft modes associated with $\ket{E}$.  This structure resembles that 
of the GHZ state in that the correlation between the hard and soft 
modes as well as that between the hard modes and early radiation 
are classical, although the correlation between the soft modes 
and early radiation is generally quantum mechanical as required 
by unitarity.  This implies that mirror modes needed for a theory 
of the interior consist of both soft modes and early radiation.

When viewed from a distance, the physics of an evaporating black hole 
is not so mysterious after all.  Scattering of high energy particles, 
or gravitational collapse, forms a bound state with a high density 
of states---a black hole---which decays into asymptotic Hawking 
quanta.  While microstates of the bound state at an intermediate 
stage cannot be resolved from the asymptotic region due to a large 
gravitational redshift, the standard rules of thermodynamics are 
obeyed~\cite{Bardeen:1973gs,Bekenstein:1974ax} throughout the 
process.  A mystery arises (only) if we consider the interior 
spacetime.  The origin of the mystery is, again, the large redshift. 
When viewed from a distance, an observer falling into a black hole 
is absorbed into the hot, stretched horizon, and yet we expect the 
existence of a description in which the observer falls smoothly 
inside the horizon, at least until it approaches the singularity. 
We have seen that such a description can be obtained in an effective 
theory erected at a fixed time, obtained after coarse-graining 
microscopic degrees of freedom that cannot be discriminated 
within the timescale of $t_{\rm H}$.  This effective theory is 
{\it intrinsically} semiclassical---the description of a falling 
object is unitary only until it hits the singularity (or escapes 
the spacetime region to which the effective theory is applicable). 
This suggests that it is meaningless to ask what happens to a 
fallen object ``after it hits the singularity'' within the interior 
picture.  The only fundamental description, which can be unitary 
for an arbitrarily long time, is that in a distant frame.  In the 
context of holography, this corresponds to the holographic, boundary 
description of the system.

The results presented in this paper provide a picture of how the 
peculiarities of a black hole emerge in a holographic boundary 
theory.  A black hole in the semiclassical description corresponds 
to a high density of states having energetically and spatially 
similar profiles in a code subspace.  As a result of this degeneracy, 
arising from a large gravitational redshift, a vast majority of 
the degrees of freedom associated with these states cannot be 
represented as those allowing for subsystem recovery.%
\footnote{In the boundary picture, this occurs because the 
 entanglement structure of a boundary state becomes such that 
 when the boundary is pulled in to the stretched horizon by 
 coarse-graining short range correlations~\cite{Nomura:2018kji}, 
 the resulting boundary state in the effective Hilbert space 
 becomes a generic, maximally entropic state.  This prevents 
 us from erecting a code subspace with subsystem recovery, since 
 the state does not have a resource of ``unentanglement'' needed 
 to have such a space~\cite{Nomura:2017fyh}.}
Only a tiny fraction---the hard modes---allow for such recovery, 
having direct association with spacetime.  The degrees of freedom 
that do not admit subsystem recovery---the soft modes---are 
indistinguishable within a characteristic timescale for the 
boundary evolution of these states; discriminating them requires 
much longer time in which the degeneracy is resolved by evolution 
involving Hawking evaporation.  However, precisely because of 
this indistinguishability, we can have a new coarse-grained 
description applicable in the characteristic timescale.  Such 
a description is useful for knowing the fate of a small object 
falling into the black hole.  Because of the large redshift, 
the applicability of this effective description is limited to 
a timescale of order the string/cutoff scale (multiplied by a 
logarithmic factor) locally near the stretched horizon; however, 
due to an extremely large boost, this corresponds to a macroscopic 
timescale perceived by the object.  The effective description 
is possible because after the coarse-graining, collective 
excitations of the soft modes (and the degrees of freedom 
entangling with them, including early Hawking radiation) appear 
as the mirror of the hard modes, with the entanglement between 
the mirror and the original hard modes taking a thermofield 
double form.  This allows for reinterpreting these collective 
modes as representing degrees of freedom that admit subsystem 
recovery in a code subspace erected on the extended, second 
boundary, which is isomorphic to the original boundary renormalized 
down to the edge of the zone, a la Ref.~\cite{Nomura:2018kji}. 
In the bulk, this gives the two-sided black hole picture applicable 
within the domain of dependence of the union of the zone and 
its mirror regions.

The physics of a black hole has a parallel in Minkowski space:\ 
an accelerating detector measuring Unruh radiation.  As in the case 
of a young black hole, this introduces entanglement between the 
detector and radiation degrees of freedom.  From the point of view 
of an inertial observer, this is entanglement between the detector 
and particles emitted from it.  This entanglement can be retrieved 
in Rindler space if we decelerate the detector adiabatically. 
This can be viewed as an analogue of information retrieval from 
a black hole, although the retrieved information in this case 
is not thermalized because of the infinite entropy of the 
Minkowski vacuum.

The discussion described above suggests that we may view the black 
hole interior as a sort of ``compactified (half) Minkowski space.'' 
As viewed from the exterior, the finiteness of the system leads 
to thermalization/scrambling of information retrieved from a black 
hole.  From the viewpoint of the interior, it is reflected in 
the fact that the description is fundamentally non-unitary, a 
manifestation of which is the existence of the singularity.  It 
would be interesting to see if other singularities in general 
relativity could be understood in similar manners.

\section*{Acknowledgments}

I thank Kanato Goto, Masahiro Nozaki, Pratik Rath, Nico Salzetta, 
Fabio Sanches, and Sean Weinberg for discussion on this and related 
topics.  This work was supported in part by the Department of Energy, 
Office of Science, Office of High Energy Physics under contract 
DE-AC02-05CH11231 and award DE-SC0019380, by the National Science 
Foundation under grant PHY-1521446, and by MEXT KAKENHI Grant 
Number 15H05895.

\end{document}